\documentclass[preprintnumbers,eqsecnum,nofootinbib,prd,twocolumn]{revtex4}  

\usepackage{graphicx}
\usepackage{latexsym}
\usepackage{amsmath}
\usepackage{amssymb}

\def\beq{\begin{equation}}
\def\eeq{\end{equation}}

\begin{document}

\title{Spin-orbit precession  along eccentric orbits: improving the knowledge of  self-force corrections and of their  effective-one-body counterparts}

\author{Donato Bini$^1$, Thibault Damour$^2$, Andrea Geralico$^1$}
  \affiliation{
$^1$Istituto per le Applicazioni del Calcolo ``M. Picone,'' CNR, I-00185 Rome, Italy\\
$^2$Institut des Hautes \'Etudes Scientifiques, 91440 Bures-sur-Yvette , France.
}

\begin{abstract}
The (first-order) gravitational self-force correction to the spin-orbit precession of a spinning compact body along a slightly
 eccentric orbit around a Schwarzschild black hole is computed through the ninth post-Newtonian order, improving recent
  results by Kavanagh et al. [Phys.\ Rev.\ D {\bf 96}, 064012 (2017).]
This information is then converted into its corresponding Effective-One-Body counterpart, thereby determining several new
post-Newtonian terms in the gyrogravitomagnetic ratio $g_{S*}$.
\end{abstract}


\maketitle

\section{Introduction}

In the newly born gravitational wave (GW) era \cite{Abbott:2016blz,Abbott:2016nmj,Abbott:2017oio,TheLIGOScientific:2017qsa}, it will become more and more important to extract accurate physical information from experimental data as
rapidly as possible. This implies constantly improving the mathematical modelling of the dynamics, and of the 
gravitational-wave emission, of inspiralling and coalescing binary systems. One of the current key methods used
in the LIGO-Virgo data analysis pipelines, is the Effective-One-Body (EOB) formalism \cite{Buonanno:1998gg,Buonanno:2000ef,Damour:2000we,Damour:2001tu,Damour:2008gu}. 
The EOB formalism is used both in the construction of hundreds of thousands
of semi-analytical templates~\cite{Taracchini:2013rva,Purrer:2015tud,Bohe:2016gbl}, describing the complete waveform emitted
by coalescing binary black holes, and in the construction of hybrid EOB-numerical waveforms that are then used
to calibrate frequency-domain phenomenological waveforms \cite{Khan:2015jqa}.

The EOB approach is based, among other building blocks, on the definition of an analytical, resummed  Hamiltonian which
 allows one to describe the coalescence process up to the merger of the two considered bodies. It is useful both
 for binary black hole systems, and for systems comprising neutron stars \cite{Damour:2009wj,Bernuzzi:2014owa,Steinhoff:2016rfi}. In recent years the necessity of making EOB theory more efficient has been driving research in 
 several analytical directions which can potentially improve the accuracy of the EOB dynamics. In particular, new knowledge
 acquired through Post-Newtonian (PN) theory (valid in the weak-field and slow motion regime),  gravitational self-force (SF) theory (valid when the mass ratio of the two bodies is very small),  Post-Minkowskian (PM) theory (valid in  the weak field regime), and numerical relativity (NR), has been usefully transcribed in terms of the basic potentials entering the
 EOB Hamiltonian. For instance, the current, fourth post-Newtonian (4PN) knowledge \cite{Damour:2014jta,Jaranowski:2015lha,Damour:2016abl,Marchand:2017pir} has been translated in EOB terms in Ref. \cite{Damour:2015isa}.
 For examples of the translation of high-PN-order SF knowledge into EOB counterparts, see, e.g., 
 Refs. \cite{Bini:2013rfa,Bini:2014ica,Hopper:2015icj,Kavanagh:2017wot}.

The aim of the present paper is to improve  the current analytical 
knowledge of {\it eccentricity-dependent} contributions to the spin-orbit precession of a spinning compact body 
orbiting a nonspinning black hole, and to translate this knowledge within the EOB formalism.
The computation of gauge-invariant, eccentricity-dependent SF effects in the spin-orbit precession of a small spinning body
was initiated in a recent paper by Akcay, Dempsey and Dolan \cite{Akcay:2016dku}. Then
 Kavanagh et al. \cite{Kavanagh:2017wot} analytically computed the PN expansion of the self-force correction to the spin-orbit precession, up to the sixth PN order and transcribed this information into the corresponding knowledge of the PN
 expansion of  the (phase-space-dependent) EOB gyrogravitomagnetic ratio $g_{S*}(r,p_r,p_{\phi})$ up to the fourth PN order in the coefficient of the square of the radial momentum, i.e. $O(u^4 p_r^2)$ included (where $u=GM/(c^2 r)$).
 
Here we shall extend the work done in Ref. \cite{Kavanagh:2017wot} to the ninth PN level for the spin precession, at the second order in eccentricity, almost doubling the number of the analytically known terms 
(because of the presence of many half-PN-order contributions).  
We shall then explicitly derive the relationship between the spin precession invariant along eccentric orbits, and the various potentials parametrizing spin-orbit effects within the EOB formalism, thereby determining the
PN expansion of the $O(p_r^2)$ contribution to the gyrogravitomagnetic ratio $g_{S*}(r,p_r,p_{\phi})$
up to the fractional seventh PN accuracy (i.e. an improvement by six half-PN-order contributions).

To make the paper self consistent we will start by briefly recalling the main computational steps of  Refs. \cite{Akcay:2016dku} and \cite{Kavanagh:2017wot}. Most of the technical details 
will, however, be relegated to an appendix. Unless differently specified we will use units so that $c=G=1$.

\section{First-order SF spin-precession invariant $\Delta \psi(u_p,e)$}

In this section we recall the basic theory underlying the derivation of the spin precession invariant
$\psi(m_2\Omega_r, m_2\Omega_\phi; m_1/m_2)$, and its first-order SF contribution $\Delta \psi$. 
Consider a binary system consisting of a spinning compact body (of mass $m_1$ and spin $S_1$) and a Schwarzschild black hole (of mass $m_2$ and spinless, $S_2=0$)  with $q \equiv \frac{m_1}{m_2} \ll 1$.
Through $O(m_1)$, the small body can be considered as following an eccentric geodesic orbit in a (regularized) perturbed spacetime $g^{\rm R}_{\alpha\beta}$, while its associated spin vector is parallelly-transported in $g^{\rm R}_{\alpha\beta}$. Here we consider the small-spin regime $|S_1|/(cGm_1^2)\ll 1$, i.e. we work linearly in $S_1$.
The regularized perturbed metric $g^{\rm R}_{\alpha\beta}$ is decomposed as
\beq
g^{\rm R}_{\alpha\beta}=\bar g_{\alpha\beta}+q \, h^{\rm R}_{\alpha\beta} + O(q^2)\,,
\eeq
where $\bar g_{\alpha\beta}$ is the background spacetime 
\begin{eqnarray}
d\bar s^2 &=&\bar g_{\alpha\beta}dx^\alpha dx^\beta\nonumber\\
&=& -fdt^2+\frac{1}{f}dr^2+r^2(d\theta^2+\sin^2\theta d\phi^2)\,, 
\end{eqnarray}
with $f=1-\frac{2m_2}{r}$, and where $q \, h^{\rm R}_{\alpha\beta}$  is the first-order SF metric perturbation. Henceforth, we shall omit the superscript R.
Let us denote by $\Omega_r=2\pi/T_r$  and $\Omega_{\phi}=\Phi/T_r$  the radial and (averaged) azimuthal angular frequencies, respectively. Here, $\Phi$ denotes the accumulated azimuthal angle from periapsis to periapsis. 
The spin precession is conveniently measured by the dimensionless quantity 
\beq
\label{psi_def}
\psi(m_2\Omega_r, m_2\Omega_\phi; q)=1-\frac{\Psi}{\Phi} 
\,,
\eeq
defined by the ratio of the amount of precession angle $\Psi$ (with respect to a polar-type basis)
accumulated by the spin vector over one radial period $T_r$, 
to the accumulated periastron precession angle $\Phi$. 

Akcay, Dempsey and Dolan \cite{Akcay:2016dku} showed how to calculate the $O(q)$, SF contribution to 
the gauge-invariant function $\psi(m_2\Omega_r, m_2\Omega_\phi; q)$, i.e. (taking into account that
$\Phi$ is the same for the perturbed, $q \ne 0$, and background, $q=0$, orbits),
\begin{eqnarray}
\Delta \psi&=& \frac1q \left[\psi(m_2\Omega_r,m_2\Omega_{\phi};q)-\psi(m_2\Omega_r,m_2\Omega_{\phi};0) \right]\nonumber\\
&=&-\frac{\Delta\Psi}{\Phi}\,,
\end{eqnarray}
with 
\beq
	\Delta \Psi=\frac1q \left[ \Psi(m_2\Omega_r,m_2\Omega_{\phi};q)-\Psi(m\Omega_r,m_2\Omega_{\phi},0) \right]\,.
\eeq
See Ref. \cite{Akcay:2016dku}, and the Appendix below, for the procedure needed to compute $\Delta \Psi$ for fixed values
of the two frequencies $(\Omega_r,\Omega_{\phi})$. After the computation of the function 
$\Delta \psi(\Omega_r,\Omega_{\phi})$, one can reexpress it as a function of the inverse 
semi latus rectum $u_p=1/p$, and eccentricity $e$, of the unperturbed orbit.

Kavanagh et al. \cite{Kavanagh:2017wot} have recently calculated,  following the approach of Ref. \cite{Akcay:2016dku}, the spin-precession invariant $\Delta \psi(u_p, e)$ up to order $O(e^2)$ in a small-eccentricity expansion, $e\ll1$, and 
up to  order $O(u_p^6)$ in the PN expansion, $u_p=1/p\ll1$.
Their calculation was based on a computation (via the Teukolsky formalism) of the PN-expanded  metric perturbation in the 
radiation gauge.
We closely follow their analysis, extending the calculation of $\Delta \psi$ up to the order $O(u_p^9)$ included.
Our final result for the spin precession invariant $\Delta \psi(u_p,e)$ reads
\begin{eqnarray} 
\label{e-exppsi}
\Delta \psi(u_p,e) &=&\Delta \psi^{(0)}(u_p)+e^2 \Delta \psi^{(2)}(u_p)\nonumber\\
&& +e^4 \Delta \psi^{(4)}(u_p)
+\mathcal{O}(e^6)\,,
\end{eqnarray}
where the PN structure of $\Delta \psi^{(2)}(u_p)$ is (note the half-PN-order terms $c_{ k+\frac12 }, c_{ k+\frac12 }^{\ln{}},\ldots$)
\begin{eqnarray}
\Delta \psi^{(2)}(u_p) &=& 
\sum_{k\ge 2}c_k^{\rm c} u_p^k+\ln u_p\,\sum_{k\ge 4}c_k^{\rm \ln{}} u_p^k\nonumber\\
&&+\sum_{k\ge 5 }c_{ k+\frac12 } u_p^{k+\frac12}+\ln^2 u_p\sum_{k\ge 7} c_k^{\ln^2{}} u_p^k\nonumber\\
&&  +\ln u_p \sum_{k\ge 8 }c_{ k+\frac12 }^{\ln{}} u_p^{k+\frac12} + \ldots\,, 
\end{eqnarray}
and explicitly
\begin{widetext}
\begin{eqnarray}
\label{result}
\Delta \psi^{(2)}(u_p) &=& 
u_p^2-\left(\frac{123}{256} \pi^2-\frac{341}{16}\right)u_p^3\nonumber\\
&-& \left(\frac{164123}{480}-\frac{536}{5} \gamma-\frac{268}{5} \ln(u_p)+\frac{23729}{4096} \pi^2+\frac{10206}{5} \ln(3)-\frac{11720}{3} \ln(2)\right)u_p^4\nonumber\\
&-&\left(\frac{4836254}{105} \ln(2)-\frac{21333485}{49152} \pi^2-\frac{9765625}{1344} \ln(5)+\frac{22682}{15} \gamma+\frac{89576921}{57600}+\frac{11341}{15} \ln(u_p)\right.\nonumber\\
&&\left.
-\frac{4430133}{320} \ln(3)\right) u_p^5
+\frac{319609}{630} \pi u_p^{11/2}\nonumber\\
&-&\left(\frac{17193359375}{145152} \ln(5)-\frac{32088966503}{2359296} \pi^2+\frac{2508913}{1890} \ln(u_p)-\frac{273329813}{945} \ln(2)\right. \nonumber\\
&&\left.+\frac{159335343}{8960} \ln(3)+\frac{146026515}{1048576} \pi^4+\frac{464068669129}{5080320}+\frac{2508913}{945} \gamma\right) u_p^6\nonumber
\end{eqnarray}
\begin{eqnarray}
&-& \frac{1586616631}{235200} \pi u_p^{13/2}\nonumber\\
&-&\left(\frac{39826256}{315} \ln(u_p) \ln(2)+\frac{2888955324477314921}{2347107840000}-\frac{1075057978433}{9702000} \ln(u_p)-\frac{678223072849}{6082560} \ln(7)\right.\nonumber\\
&&+\frac{79652512}{315} \gamma \ln(2)-\frac{15912612}{175} \ln(2) \ln(3)+\frac{7219504}{525} \gamma \ln(u_p)-\frac{2411543359375}{2838528} \ln(5)\nonumber\\
&&-\frac{7956306}{175} \ln(3)^2-\frac{2387982140729}{8731800} \ln(2)-\frac{1075057978433}{4851000} \gamma-\frac{465082867177871}{6606028800} \pi^2\nonumber\\
&&+\frac{1804876}{525} \ln(u_p)^2-\frac{15912612}{175} \gamma \ln(3)+\frac{299782486660473}{275968000} \ln(3)+\frac{7219504}{525} \gamma^2\nonumber\\
&&\left.
-\frac{314165501411}{335544320} \pi^4+\frac{80263696}{175} \ln(2)^2-\frac{134944}{5}\zeta(3)-\frac{7956306}{175} \ln(u_p) \ln(3)\right) u_p^7\nonumber\\
&+& \frac{2404331748779}{279417600} \pi u_p^{15/2}\nonumber\\
&-&\left(\frac{3173828125}{7056} \ln(2) \ln(5)-\frac{19171249336}{11025}\ln(u_p) \ln(2)+\frac{3173828125}{7056}\gamma \ln(5)+\frac{621149553}{784}\gamma \ln(3)\right.\nonumber\\
&&-\frac{12590844685671737819611}{939781979136000}+\frac{33310259864964463}{443925135360}\pi^2+\frac{3972491619599291}{5297292000}\ln(u_p)\nonumber\\
&&-\frac{812331139710343959}{100452352000}\ln(3)+\frac{3173828125}{14112}\ln(u_p) \ln(5)-\frac{38342498672}{11025}\gamma \ln(2)+\frac{5551264}{21}\zeta(3)\nonumber\\
&&+\frac{55100101995388051}{23721984000}\ln(7)+\frac{621149553}{784}\ln(2) \ln(3)+\frac{3173828125}{14112}\ln(5)^2+\frac{88630614687481099}{7945938000}\ln(2)\nonumber\\
&&+\frac{3978068608616891}{2648646000}\gamma-\frac{371280152}{2205}\gamma \ln(u_p)+\frac{621149553}{1568}\ln(u_p) \ln(3)+\frac{621149553}{1568}\ln(3)^2-\frac{371280152}{2205}\gamma^2\nonumber\\
&&\left.+\frac{196313675703125}{21697708032}\ln(5)+\frac{31118085613898053}{257698037760}\pi^4-\frac{7893208952}{1225}\ln(2)^2-\frac{92820038}{2205}\ln(u_p)^2\right) u_p^8\nonumber\\
&-&\left(-\frac{21157957560083191439}{33563642112000}\pi+\frac{166628746}{315}\pi \ln(2)+\frac{6610847233}{165375}\pi \ln(u_p)+\frac{13221694466}{165375}\pi \gamma\right.\nonumber\\
&&\left. 
-\frac{926441631}{6125}\pi \ln(3)-\frac{123567238}{4725}\pi^3\right) u_p^{17/2}\nonumber\\
&-&\left(\frac{89957972}{315}\zeta(3)-\frac{113231016807891871}{5175705600} \ln(7)+\frac{47576994975313901}{21189168000} \gamma+\frac{72839234785800578877}{4419903488000} \ln(3)\right.\nonumber\\
&&+\frac{1362682591}{19845} \ln(u_p)^2+\frac{1590291075741430578125}{33141699182592} \ln(5)-\frac{27467645056526844769}{419545526400} \ln(2)\nonumber\\
&&-\frac{962681186487}{268435456} \pi^6+\frac{5450730364}{19845} \gamma \ln(u_p)+\frac{11600054253556}{1091475} \ln(u_p) \ln(2)+\frac{23200108507112}{1091475} \gamma \ln(2)\nonumber\\
&&-\frac{873170066079}{1724800} \ln(u_p) \ln(3)+\frac{815583044961}{862400} \ln(2) \ln(3)-\frac{873170066079}{862400} \gamma \ln(3)-\frac{28312548828125}{3592512} \gamma \ln(5)\nonumber\\
&&-\frac{28312548828125}{3592512} \ln(2) \ln(5)-\frac{28312548828125}{7185024} \ln(u_p) \ln(5)-\frac{873170066079}{1724800} \ln(3)^2\nonumber\\
&&+\frac{399034285145396}{9823275} \ln(2)^2+\frac{5450730364}{19845} \gamma^2+\frac{47354559565317101}{42378336000} \ln(u_p)\nonumber\\
&&-\frac{1632714242298331008005890223}{2638907797413888000}-\frac{28312548828125}{7185024} \ln(5)^2-\frac{517077970581977291}{2959500902400} \pi^2\nonumber\\
&&\left.+\frac{411339398981140702321}{65970697666560} \pi^4\right)u_p^9+O_{\ln{}}(u_p^{19/2})\,.
\end{eqnarray}
\end{widetext}
When comparing with the lower-accuracy result of  Ref. \cite{Kavanagh:2017wot}, one should note that the whole 
 $O(u_p^5)$ term  was misprinted there (as being simply exactly the same as the $O(u_p^4)$ term).
The terms from $u_p^{13/2}$ (included) up to $u_p^{9}$ (included) are new with this work, and represent  one of the main outcomes of the present paper.

The zero-eccentricity term $\Delta \psi^{(0)}(u_p)$ in Eq. \eqref{e-exppsi} is related, as shown in \cite{Akcay:2016dku}, 
to the spin precession invariant $\Delta\psi_{\rm (circ)}(u_p)$ directly computed along circular orbits \cite{Bini:2014ica,Bini:2015mza,Shah:2015nva,Kavanagh:2015lva} via
\begin{eqnarray} \label{dpsi0}
\Delta \psi^{(0)}(u_p) &=&
\Delta\psi_{\rm (circ)}(u_p)\\
&+& \frac14 \frac{(1-3u_p)^{1/2}(1-6 u_p)}{1-\frac{39}{4}u_p+\frac{43}{2}u_p^2}(\rho(u_p)-4u_p) \,,\nonumber
\end{eqnarray}
where $\rho(u_p)$ is the EOB function measuring the periastron precession at the 1SF-level \cite{Damour:2009sm,Bini:2016qtx}.
[Note that the expression for $\Delta\psi_{\rm (circ)}(y)$ given in Ref. \cite{Shah:2015nva} is
incorrect for the fractional power terms $y^{19/2}$ and beyond.]

The higher-order-in-eccentricity contributions to $\Delta \psi(u_p,e)$, starting from $e^4\Delta \psi^{(4)}(u_p)$, 
present an analytical challenge that we leave to future work. The present knowledge of $\Delta \psi(u_p,e)$ beyond the $O(e^2)$ level consists of the lowest-PN-order  $O(e^4)$ term (derived in Ref. \cite{Akcay:2016dku}), namely
\beq
\label{contr_e4}
\Delta \psi^{(4)}(u_p)=- \frac12 u_p^3\,.
\eeq

\section{Improving the knowledge of the EOB  gyrogravitomagnetic ratio $g_{S*}(u, p_r,p_\phi)$}

In EOB theory, the total Hamiltonian of a two-body system  is expressed in terms of the  ``effective EOB Hamiltonian",  $H_{\rm eff}$, via
\begin{eqnarray}
\label{H_Heff}
H({\mathbf R},{\mathbf P},{\mathbf S}_1,{\mathbf S}_2)&=&Mc^2\sqrt{1+2\nu \left(\frac{H_{\rm eff}}{\mu c^2}-1\right)}\nonumber\\
&\equiv& Mc^2 h\,,
\end{eqnarray}
where 
\begin{eqnarray}
&& M=m_1+m_2\,,\qquad \mu=\frac{m_1m_2}{m_1+m_2}\,,\nonumber\\
&& \nu=\frac{\mu}{M}=\frac{m_1m_2}{(m_1+m_2)^2}\,.
\end{eqnarray}
When considering spinning bodies, the effective EOB Hamiltonian  $H_{\rm eff}$ is decomposed into the sum
of  an orbital part and a spin-orbit part
\beq
H_{\rm eff}=H^{\rm O}_{\rm eff}+H^{\rm SO}_{\rm eff}\,.
\eeq
Here, we work linearly in the spins, so that the orbital part  $H^{\rm O}_{\rm eff}$ will be independent of the spins,
while the spin-orbit part will be linear in the spins. The structure of  the {\it orbital} part is
\beq
\label{orb_ham}
H^{\rm O}_{\rm eff}=c^2 \sqrt{A \left(\mu^2c^2 +{\mathbf P}^2+\left(\frac{1}{B}-1\right)P_R^2+ Q\right)}\,,
\eeq
where
\beq
{\mathbf P}^2=\frac{P_R^2}{B}+\frac{{\mathbf L}^2}{R^2}=\frac{P_R^2}{B}+\frac{P_\phi^2}{R^2}\,.
\eeq
Here  ${\mathbf L}={\mathbf R}\times {\mathbf P}$ denotes the orbital angular momentum ($|{\mathbf L}|\equiv P_\phi$), $A(R)$ and $B(R)$ are the two main EOB radial potentials
and the phase-space extra potential $Q(R,P_R)$ is at least quartic in the radial momentum $P_R$.
The  structure of the spin-orbit  part of the effective Hamiltonian is 
\begin{eqnarray}
H^{\rm SO}_{\rm eff}&=& G_S^{\rm phys}(R, P_R^2, {\mathbf L}^2) {\mathbf L}\cdot {\mathbf S}\nonumber\\
&&+G_{S*}^{\rm phys}(R, P_R^2, {\mathbf L}^2)  {\mathbf L}\cdot {\mathbf S}_*  \,.
\end{eqnarray}
It involves the following two symmetric combination of the spin vectors ${\mathbf S}_1$ and ${\mathbf S}_2$ of the system
\beq
{\mathbf S}\equiv{\mathbf S}_1+{\mathbf S}_2\,,\qquad {\mathbf S}_*\equiv\frac{m_2}{m_1} {\mathbf S}_1+\frac{m_1}{m_2}{\mathbf S}_2\,.
\eeq
In the parallel-spin case ${\mathbf L}\cdot {\mathbf S}=LS=P_\phi S$ and ${\mathbf L}\cdot {\mathbf S}_*=LS_*=P_\phi S_*$.
It is convenient to work with  the following dimensionless  variables 
\begin{eqnarray}
\label{rescal}
&& r=\frac{c^2 R}{GM}\,,\quad u=\frac{GM}{c^2 R}\equiv\frac{1}{r}\,,\quad j\equiv p_\phi=\frac{c P_\phi}{GM\mu}\,,\nonumber\\
&& p_r=\frac{P_R}{\mu c}\,, \quad \hat Q= \frac{Q}{\mu^2 c^2} \,,
\end{eqnarray}
as well as
\beq
\label{rescal2}
\hat H_{\rm eff}=\frac{H_{\rm eff}}{\mu c^2}\equiv \hat H^{\rm O}_{\rm eff}+\hat H^{\rm SO}_{\rm eff}\,.
\eeq
The Finslerlike contribution $\hat Q$ has the structure
\beq
\hat Q= \nu \sum_{n=2}^\infty q_{2n}(u)p_r^{2n} + O(\nu^2)\,.
\eeq
As we work linearly in the spins, we can replace the dimensionfull spin-orbit coupling functions,
$G_S^{\rm phys}$ and $G_{S*}^{\rm phys}$, entering $H^{\rm SO}_{\rm eff}$ by the corresponding dimensionless gyrogravitomagnetic ratios $g_S$ and $g_{S*}$ defined as
\begin{eqnarray}
g_S(u,p_r,p_\phi) &=& R^3 G_S^{\rm phys}\,,\nonumber\\
g_{S*}(u,p_r,p_\phi)&=& R^3 G_{S*}^{\rm phys}\,.
\end{eqnarray}
Here, we shall parametrize the SF expansions (i.e. expansions in powers of $\nu$) of $g_S$ and $g_{S*}$ as
\begin{eqnarray} \label{nuexpgS}
g_S(u,p_r,p_\phi; \nu)&=& 2 + O(\nu) \,, \nonumber \\
g_{S*}(u,p_r,p_\phi; \nu)&=&g_{S*}^{(\nu^0)}(u,p_r,p_\phi)+\nu g_{S*}^{(\nu^1)}(u,p_r)\nonumber\\
&& +\nu^2 g_{S*}^{(\nu^2)}(u,p_r)+\ldots
\end{eqnarray}
with the test-mass limit of $g_{S*}(u,p_r,p_\phi)$ written in the form \cite{Barausse:2009aa,Bini:2015xua}
\beq \label{gSs0}
g_{S*}^{(\nu^0)}(u,p_r,p_\phi)=\frac{2 }{1+ \frac{1}{\sqrt{1-2 u}}  }\, \frac{1}{\widetilde R}+\frac{1}{1+ \widetilde R}\,,
\eeq
where
\beq
\widetilde R(u,p_r,p_\phi)=[1+p_\phi^2 u^2+(1-2u)p_r^2]^{1/2}\,.
\eeq
In the SF expansion of $g_{S*}(u,p_r,p_\phi; \nu)$ (second equation in Eq. \eqref{nuexpgS}), we have made
a specific gauge-choice for the phase-space dependence of the SF contributions: namely, following the spirit of
Ref. \cite{Damour:2008qf}, we have represented them as functions of $u$ and $p_r$, without allowing for a
dependence on $p_\phi$.
The first-order self-force (1SF) contribution to $g_{S*}$ can then be expanded in (even) powers of the radial momentum:
\begin{eqnarray} 
\label{gSs1SF}
g_{S*}^{(\nu^1)}(u,p_r)&=&g_{S*}^{\rm 1SF0}(u)+g_{S*}^{\rm 1SF2}(u)p_r^2\nonumber\\
&+& g_{S*}^{\rm 1SF4}(u)p_r^4+g_{S*}^{\rm 1SF6}(u)p_r^6+\ldots\,.
\end{eqnarray}
In turn, the various coefficients $g_{S*}^{\rm 1SF0}(u)$, $g_{S*}^{\rm 1SF2}(u)$, etc. of this $p_r^2$ expansion, have PN expansions in $u$ which start as
\begin{eqnarray}
g_{S*}^{\rm 1SF0}(u)&=& -\frac34 u-\frac{39}{4} u^2+\left(-\frac{7987}{192}+\frac{41}{32}\pi^2\right) u^3\nonumber\\
&& +O(u^4) \,,\nonumber\\
g_{S*}^{\rm 1SF2}(u)&=& -\frac{9}{4} -\frac{9}{4} u-\frac{717}{32} u^2 +O(u^3) \,,\nonumber\\
g_{S*}^{\rm 1SF4}(u)&=&  \frac{5}{2}  +O(u)\,.
\end{eqnarray}
Let us introduce the following notation for the coefficients of the various powers of $u$ in the PN expansion of 
$g_{S*}^{\rm 1SFn}(u)$
\begin{eqnarray} 
\label{PNexpgSsn}
g_{S*}^{\rm 1SFn}(u) &=& 
\sum_{k\ge 1}g_{*nk}^{\rm c} u^k+\ln u\,\sum_{k\ge 3}g_{*nk}^{\rm \ln{}} u^k\nonumber\\
&+& \sum_{k\ge 4}g_{*n\left(k+\frac12\right) } u^{k+\frac12}  +\ln^2 u\sum_{k\ge 6} g_{*nk}^{\ln^2{}} u^k \nonumber\\
&+& \ln u \sum_{k\ge 7}g_{*n\left(k+\frac12\right) } u^{k+\frac12} \ldots\,. 
\end{eqnarray}
Only a few of these coefficients were determined in Ref. \cite{Kavanagh:2017wot}, namely:
\begin{eqnarray}
g_{*22}&=& -\frac{717}{32} \nonumber\\
g_{*23}^c&=& \frac{1447441}{960}-\frac{4829}{256}\pi^2-\frac{16038}{5}\ln(3)\nonumber\\
&& +\frac{46976}{15}\ln(2)-\frac{512}{5}\gamma\nonumber\\
g_{*23}^{\ln{}}&=&-\frac{256}{5} \nonumber\\
g_{*24}^c&=&- \left[\frac{184655453}{38400}\right]_{\rm corrected}+\frac{19162}{35}\gamma+\frac{2097479}{8192}\pi^2\nonumber\\
&& +\frac{454167}{20}\ln(3)-\frac{1081966}{35}\ln(2) \nonumber\\
g_{*24}^{\ln{}}&=&+\frac{9581}{35} \,.
\end{eqnarray}
Note that the first (rational) term in  $g_{*24}^c$ was misprinted in Ref. \cite{Kavanagh:2017wot} as
\beq
- \left[\frac{185195453}{38400}\right]_{\rm incorrect} \,.
\eeq
In the present work, we have derived (by using the relation between the PN expansion \eqref{PNexpgSsn}
and the PN expansion of $\Delta \psi^{(2)}$) additional terms in the PN expansion of $g_{S*}^{\rm 1SF2}(u)$, namely:
\begin{widetext}
\begin{eqnarray}
g_{*24.5} &=& -\frac{714653}{3150}\pi
\,,\nonumber\\
g_{*25}^c &=&\frac{1136089}{210}\gamma-\frac{19531250}{567}\ln(5)-\frac{82368981}{560}\ln(3)+\frac{121254173}{378}\ln(2)+\frac{1322959637}{196608}\pi^2+\frac{21119805}{524288}\pi^4\nonumber\\
&&
-\frac{2157036100969787}{26604864000}
\,,\nonumber\\
g_{*25}^{\ln{}} &=& \frac{1136089}{420}
\,,\nonumber\\
g_{*25}^{\ln^2{}} &=& 0
\,,\nonumber\\
g_{*25.5} &=& \frac{214163953}{100800}\pi
\,,\nonumber\\
g_{*25.5}^{\ln{}} &=& 0
\,,\nonumber\\
g_{*25.5}^{\ln^2{}} &=& 0
\,,\nonumber\\
g_{*26}^c &=&-9024\zeta(3)-\frac{62597340649}{485100}\gamma+\frac{129552734375}{266112}\ln(5)-\frac{695576369871}{1232000}\ln(3)-\frac{640198452653}{661500}\ln(2)\nonumber\\
&&
-\frac{126056977883647}{4459069440}\pi^2+\frac{549062083551}{167772160}\pi^4-\frac{149101504}{1575}\gamma\ln(2)+\frac{2808108}{25}\ln(3)\ln(2)+\frac{2808108}{25}\ln(3)\gamma\nonumber\\
&&
-\frac{79378592}{525}\ln(2)^2+\frac{160928}{35}\gamma^2+\frac{1404054}{25}\ln(3)^2+\frac{357317643875067280292753}{1035737850115584000}
\,,\nonumber
\end{eqnarray}
\begin{eqnarray}
g_6^{\ln{}} &=& -\frac{74550752}{1575}\ln(2)+\frac{1404054}{25}\ln(3)+\frac{160928}{35}\gamma-\frac{62597340649}{970200}
\,,\nonumber\\
g_{*26}^{\ln^2{}} &=& \frac{40232}{35}
\,,\nonumber\\
g_{*26.5} &=&  \frac{193000744063}{93139200}\pi
\,,\nonumber\\
g_{*26.5}^{\ln{}} &=& 0
\,,\nonumber\\
g_{*26.5}^{\ln^2{}} &=& 0
\,,\nonumber\\
g_{*27}^c &=& \frac{7394848}{35}\zeta(3)-\frac{4747561509943}{15444000}\ln(7)+\frac{16112853388177387}{15891876000}\gamma-\frac{29717454296875}{8805888}\ln(5)\nonumber\\
&&
+\frac{2141915038593471}{448448000}\ln(3)+\frac{5373188171264357}{2270268000}\ln(2)+\frac{36023768285157071}{208089907200}\pi^2-\frac{9570352156443723}{10737418240}\pi^4\nonumber\\
&&
+ \frac{448804392116006600471071523}{5192955612149760000}
+\frac{45230944}{63}\gamma\ln(2)-\frac{23385591}{25}\ln(3)\ln(2)-\frac{23385591}{25}\ln(3)\gamma\nonumber\\
&&
+\frac{19919007824}{11025}\ln(2)^2-\frac{1722628912}{11025}\gamma^2-\frac{23385591}{50}\ln(3)^2
\,,\nonumber\\
g_{*27}^{\ln{}} &=&  \frac{16069729193463787}{31783752000}-\frac{1722628912}{11025}\gamma-\frac{23385591}{50}\ln(3)+\frac{22615472}{63}\ln(2)
\,,\nonumber\\
g_{*27}^{\ln^2{}} &=& -\frac{430657228}{11025}
\,.
\end{eqnarray}

Summarizing, the present, first-order self-force knowledge of $g_{S*}=g_{S*}^{\rm 1SF0}(u)+g_{S*}^{\rm 1SF2}(u)p_r^2+ g_{S*}^{\rm 1SF4}(u)p_r^4 + O(p_r^6)$ is  the following:
\begin{eqnarray}
g_{S*}^{\rm 1SF0}(u)&=& -\frac34 u-\frac{39}{4} u^2+\left(-\frac{7987}{192}+\frac{41}{32}\pi^2\right) u^3 \nonumber\\
&+& \left(-\frac{11447}{120}-48\gamma-\frac{1456}{15}\ln(2)-24\ln(u)+\frac{26943}{2048}\pi^2\right)u^4 \nonumber\\
&+&\left(-\frac{729}{7}\ln(3)+\frac{4216}{35}\ln(u)+\frac{8432}{35}\gamma+\frac{62296}{105}\ln(2)+\frac{1404359}{8192}\pi^2-\frac{487501139}{268800}\right)u^5\nonumber\\
&-& \frac{93304}{1575}\pi u^{11/2}\nonumber\\
&+& \left(\frac{447572}{2835}\ln(u)+\frac{895144}{2835}\gamma-\frac{1937576}{2835}\ln(2)+\frac{37179}{35}\ln(3)+\frac{16790137}{1048576}\pi^4-\frac{1780964579}{7077888}\pi^2\right.\nonumber\\
&& \left. -\frac{15354135144661823}{11732745024000}\right) u^6\nonumber\\
&+& \frac{3928339}{12600}\pi u^{13/2}\nonumber\\
&+& \left(\frac{14552}{105}\ln(u)^2-\frac{197060521717}{335544320}\pi^4-\frac{30895906513}{8731800}\ln(u)-\frac{30895906513}{4365900}\gamma+\frac{499904}{225}\ln(2)\gamma \right.\nonumber\\
&& -\frac{6007689}{1760}\ln(3)+\frac{4484838768980344811772817}{53858368206010368000}+\frac{1167584}{525}\ln(2)^2-\frac{17567377338739}{8918138880}\pi^2+\frac{58208}{105}\gamma\ln(u)\nonumber\\
&& \left.-1088\zeta(3)-\frac{1953125}{3564}\ln(5)-\frac{22149706021}{3118500}\ln(2)+\frac{58208}{105}\gamma^2+\frac{249952}{225}\ln(2)\ln(u)\right)u^{7}\nonumber\\
&+& \frac{8667496367}{34927200}\pi u^{15/2}\nonumber\\
&+&\left(\frac{54784}{105}\gamma\ln(u)+\frac{219136}{105}\ln(2)^2-\frac{147087852683}{17463600}\ln(2)+\frac{54784}{105}\gamma^2+\frac{219136}{105}\ln(2)\gamma-\frac{148697438501}{30965760}\pi^2\right.\nonumber\\
&& -\frac{2203104637}{33554432}\pi^4-\frac{9765625}{38016}\ln(5)+\frac{109568}{105}\ln(2)\ln(u)-\frac{98943259619}{17463600}\gamma-\frac{98943259619}{34927200}\ln(u)\nonumber\\
&&\left. +\frac{13696}{105}\ln(u)^2 -\frac{5875497}{4928}\ln(3)+\frac{531841058042717}{9779616000}-1024\zeta(3)\right) u^{8}\nonumber\\
&+& \frac{87641066621}{209563200}\pi u^{17/2}+O_{\ln{}}(u^{9})\,,
\end{eqnarray}
\begin{eqnarray}
\label{gsstar1SF2}
g_{S*}^{\rm 1SF2}(u)&=& 
-\frac{9}{4} -\frac{9}{4} u-\frac{717}{32} u^2\nonumber\\
&+& \left(\frac{1447441}{960}-\frac{4829}{256}\pi^2-\frac{16038}{5}\ln(3)+\frac{46976}{15}\ln(2)-\frac{512}{5}\gamma-\frac{256}{5}\ln(u)\right) u^3\nonumber\\
&+& \left(-\frac{184655453}{38400}+\frac{19162}{35}\gamma+\frac{2097479}{8192}\pi^2+\frac{454167}{20}\ln(3)-\frac{1081966}{35}\ln(2)+\frac{9581}{35}\ln(u)\right) u^4
\nonumber\\
&-& \frac{714653}{3150}\pi u^{9/2}\nonumber\\
&+& \left(\frac{1136089}{210}\gamma-\frac{19531250}{567}\ln(5)-\frac{82368981}{560}\ln(3)+\frac{121254173}{378}\ln(2)+\frac{1322959637}{196608}\pi^2+\frac{21119805}{524288}\pi^4\right.\nonumber\\
&& \left.
-\frac{2157036100969787}{26604864000}+\frac{1136089}{420}\ln(u)\right) u^5\nonumber\\
&+& \frac{214163953}{100800}\pi u^{11/2} \nonumber\\
&+& \left[
-9024\zeta(3)-\frac{62597340649}{485100}\gamma+\frac{129552734375}{266112}\ln(5)-\frac{695576369871}{1232000}\ln(3)-\frac{640198452653}{661500}\ln(2)\right.\nonumber\\
&&
-\frac{126056977883647}{4459069440}\pi^2+\frac{549062083551}{167772160}\pi^4-\frac{149101504}{1575}\gamma\ln(2)+\frac{2808108}{25}\ln(3)\ln(2)+\frac{2808108}{25}\ln(3)\gamma\nonumber\\
&&
-\frac{79378592}{525}\ln(2)^2+\frac{160928}{35}\gamma^2+\frac{1404054}{25}\ln(3)^2+\frac{357317643875067280292753}{1035737850115584000}\nonumber\\
&&\left.
+\left(-\frac{74550752}{1575}\ln(2)+\frac{1404054}{25}\ln(3)+\frac{160928}{35}\gamma-\frac{62597340649}{970200}\right)\ln(u)
+\frac{40232}{35}\ln(u)^2
\right]u^6\nonumber\\
&+& \frac{193000744063}{93139200}\pi u^{13/2}\nonumber\\
&+&\left[
\frac{7394848}{35}\zeta(3)-\frac{4747561509943}{15444000}\ln(7)+\frac{16112853388177387}{15891876000}\gamma-\frac{29717454296875}{8805888}\ln(5)\right.\nonumber\\
&&
+\frac{2141915038593471}{448448000}\ln(3)+\frac{5373188171264357}{2270268000}\ln(2)+\frac{36023768285157071}{208089907200}\pi^2-\frac{9570352156443723}{10737418240}\pi^4\nonumber\\
&&
+\frac{448804392116006600471071523}{5192955612149760000}
+\frac{45230944}{63}\gamma\ln(2)-\frac{23385591}{25}\ln(3)\ln(2)-\frac{23385591}{25}\ln(3)\gamma\nonumber\\
&&
+\frac{19919007824}{11025}\ln(2)^2-\frac{1722628912}{11025}\gamma^2-\frac{23385591}{50}\ln(3)^2\nonumber\\
&&
+\left(\frac{16069729193463787}{31783752000}-\frac{1722628912}{11025}\gamma-\frac{23385591}{50}\ln(3)+\frac{22615472}{63}\ln(2)\right)\ln(u)\nonumber\\
&&\left.
-\frac{430657228}{11025}\ln(u)^2\right] u^{7}+O_{\ln{}}(u^{7.5})\,,
\end{eqnarray}
\end{widetext}
and
\begin{eqnarray}
\label{g_Sparams}
g_{S*}^{\rm 1SF4}(u)&=& \frac52 + O(u)\,.
\end{eqnarray}
In these expressions $O_{\ln{}}(u^{n})$ denotes an error term of order $u^n$ modulo a coefficient depending on $\ln u$.

\section{Convergence of the PN-expanded spin-orbit functions, and comparison to numerical SF data}
 
As was pointed out in many previous works (notably Refs. \cite{Damour:1997ub,Bini:2014zxa,Bini:2016qtx}), the speed of
convergence of PN expansions is essentially determined by the distance between the origin and the first
expected singularity of the corresponding exact function. For instance, the fact that the 1SF contribution, 
$a_{\rm 1SF}(u)$, to the EOB $A(u ; \nu)$ potential  ($A(u; \nu) = 1-2u + \nu a_{\rm 1SF}(u) + O(\nu^2)$)
has its first singularity at the lightring (LR) \cite{Akcay:2012ea}, $a_{\rm 1SF}(u) \sim (1-3u)^{-1/2}$, suggests that 
the $n$th term in the PN expansion\footnote{We recall that $u=GM/(c^2r)$ so that (modulo the 
conventional consideration of what low-order term is considered as being ``Newtonian"), a term of order $u^n$
is of $n$PN order.} $a_{\rm 1SF}(u)$ is roughly of order $\sim (3 u)^n$, and that the remainder
after the $n$th term is roughly of order $\sim (3 u)^{(n+\frac12)}/(1-3u)$ \cite{Bini:2014ica,Bini:2014zxa,Bini:2016qtx}.
In the present work, we are mainly interested in the PN expansion of the $O(e^2)$ contribution to 
a SF function of $u_p=1/p$ and $e$. In the $p,e$ plane, the {\it separatrix} of equation $p=6+2e$ 
(see, e.g., Ref. \cite{Barack:2010tm}) marks the boundary between stable and unstable (plunging) eccentric orbits. This boundary
(with its attendant change of character of the orbit) is likely to introduce a singularity in generic dynamical
functions of $p$ and $e$. When expanding such functions in powers of $e$, this will then induce a singularity
in the $u$-dependent coefficients of this expansion at the location $p=6$, i.e. at the Last Stable (circular) Orbit (LSO),
namely, $u_p=u_{\rm LSO}=1/6$.
For instance, Eq. (5.26) of \cite{Tiec:2015cxa} shows (when using the regularity of the EOB 1SF
potentials $a_{\rm 1SF}(u)$ and $d_{\rm 1SF}(u)$ at $u=u_{\rm LSO}=1/6$) that
the term of order $e^2$, say $z_{1SF}^{(2)}(u_p)$, in the eccentricity expansion of the 1SF contribution
to the averaged redshift of particle 1, has a singularity of the form $z_{1SF}^{(2)}(u_p) \sim (1-6 u_p)^{-1}$
at $u_p=u_{\rm LSO}=1/6$.  We similarly expect $\Delta \psi^{(2)}(u_p)$ to have a singularity at $u_p=u_{\rm LSO}=1/6$. By the general argument above, this singularity should entail that the 
$n$th term in the PN expansion of $\Delta \psi^{(2)}(u_p)= \sum_n C_n^{\psi^{(2)}}\!(\ln u_p) u_p^n$
has  a value, when evaluated at $u_p=u_{\rm LSO}=1/6$, that is roughly independent of $n$, say
\begin{eqnarray} 
\label{Cnpsi2}
C_n^{\psi^{(2)}\!, \,\rm LSO} &\equiv& \left[C_n^{\psi^{(2)}}\!(\ln u_p) u_p^n\right]_{u_p=\frac16} \nonumber\\
&\sim& \pm c( \psi^{(2)})\,,\label{theoryestimate}
\end{eqnarray} 
where $c( \psi^{(2)})$ is a number of order unity. In turn, this behavior implies that the value of the $n$th PN term
at any $u_p $ is roughly of order
\beq
C_n^{\psi^{(2)}}\!(\ln u_p) u_p^n \sim C_n^{\psi^{(2)}\!, \,\rm LSO} (6u_p)^n \sim \pm c( \psi^{(2)}) (6u_p)^n\,.
\eeq
Actually, things might be more subtle than just explained. Indeed, as the SF function $\Delta \psi(u_p,e)$
comes from SF expanding the function $\psi(\Omega_r,\Omega_\phi)$ it might inherit singularities at
the other separatrix where the two frequencies $\Omega_r,\Omega_\phi$ become degenerate \cite{Barack:2011ed,Warburton:2013yj}, i.e. where
the Jacobian $J=\partial (\Omega_r,\Omega_\phi)/\partial(p,e)$ vanishes. Eq. (13) of \cite{Warburton:2013yj}
shows that this occurs when $4 p^2 - 39 p + 86= 4 p^2 (1- \frac{39}{4} u_p + \frac{43}{2} u_p^2)=0$. The relevant
root is 
\begin{eqnarray}
u_{\rm isopairing}&=&\frac1{172} (39 - \sqrt{145})\nonumber\\ 
&\simeq& 0.1567349 \simeq \frac1{6.380199} \,.
\end{eqnarray}
For instance, Eq. \eqref{dpsi0} above shows that the zero-eccentricity $\Delta \psi^{(0)}(u_p)$ limit of $\Delta \psi(u_p,e)$ has its first singularity at $u_p=u_{\rm isopairing}< u_{\rm LSO}$. However, Eq. \eqref{dpsi0} shows also
that there is an extra factor $1-6u$ in the numerator of the singular piece in $\Delta \psi^{(0)}(u_p)$, so that one
expects PN-expansion coefficients of the rough type $\sim c (u/u_{\rm isopairing})^n$ with a numerically
small prefactor $c \propto 1-6 u_{\rm isopairing} \simeq 0.05959$.

In Table \ref{tab:1}, we list the numerical values of the successive $O_{\ln}(u^n)$ contributions to
both $\Delta \psi^{(0)}(u_p)$ and $\Delta \psi^{(2)}(u_p)$, evaluated at $u=u_{\rm LSO}=1/6$. 
[In the case of $\Delta \psi^{(0)}$, we are neglecting here the (fractionally small) difference between 
$u_{\rm isopairing}$ and $\frac16$.]
The results
are compatible with the expectations just explained. In particular, the coefficients $C_n^{\psi^{(2)}\!, \,\rm LSO}$, Eq. \eqref{Cnpsi2},
seem to stabilize at values of order $\sim \pm1$ as $n$ gets large. [Note that the PN order $n$ takes, after
a while, both integer and half-integer values.]


\begin{table}[t]
  \caption{
\label{tab:1} 
The numerical values of the successive $O_{\ln}(u^n)$ contributions to
both $\Delta \psi^{(0)}(u_p)$ and $\Delta \psi^{(2)}(u_p)$, evaluated at $u=u_{\rm LSO}=1/6$.
Here, $\eta \equiv \frac1c$ counts the half-PN orders.
}
  \begin{center}
    \begin{ruledtabular}
      \begin{tabular}{|l|c|c|}
\hline 
$PN$ & $\Delta \psi^{(0)}(1/6)$ & $\Delta \psi^{(2)}(1/6)$ \cr
\hline 
$\eta^2$    & -0.166667      & - \cr 
$\eta^4$    &  0.062500      &  0.027778  \cr 
$\eta^6$    &  0.126016      & +0.076715  \cr 
$\eta^8$    & -0.004643      & +0.024805  \cr 
$\eta^{10}$ & -0.052192      & -0.233017  \cr 
$\eta^{11}$ & +0.026164      & +0.083675  \cr 
$\eta^{12}$ & +0.109184      & +.4389648  \cr 
$\eta^{13}$ & -0.034307      & -0.185439  \cr 
$\eta^{14}$ & -0.001642      & +0.085211  \cr 
$\eta^{15}$ & +0.004381      & +0.039424  \cr 
$\eta^{16}$ & -0.081211      & -0.977197  \cr 
$\eta^{17}$ & +0.047979      & +0.544810 \cr 
$\eta^{18}$ & +0.057535      & +0.907178  \cr 
$\eta^{19}$ & -0.044808      & - \cr 
$\eta^{20}$ & +0.387238$\times 10^{-3}$  & - \cr 
$\eta^{21}$ & -0.196585$\times 10^{-3}$  & - \cr 
\end{tabular}
\end{ruledtabular}
\end{center}
\end{table}

These results allow us to write down a rough theoretical estimate of the remainder of any truncated PN expansion, such as
\beq \label{psi2NPN}
\left[\Delta \psi^{(2)}(u_p)\right]^{N \,\rm PN}(u)= \sum_{n=2}^N C_n^{\psi^{(2)}}(\ln u_p) u_p^n \,.
\eeq
Namely, one expects the absolute value of the $N$-PN remainder,
$\left[\Delta \psi^{(2)}(u_p)\right]^{\rm exact} - \left[\Delta \psi^{(2)}(u_p)\right]^{N \,\rm PN}$, to be roughly of order
(using the fact that, in the cases we shall consider, the next term differs by a half PN order)
\beq \label{sigmath}
\sigma_{N \, \rm PN}^{\rm th}(\Delta \psi^{(2)}(u))= \left|C_{N+\frac12}^{\psi^{(2)}\!, \,\rm LSO}\right| \frac{(6 u)^{(N+\frac12)}}{(1-6u)^{\alpha_{N}}} \,.
\eeq
Here, like in our previous works \cite{Bini:2014zxa,Bini:2016qtx}, we allow for the possibility of having not only
an overall numerical prefactor, namely $ C_{N+\frac12}^{\psi^{(2)}\!, \,\rm LSO}$, but also to correct the contribution
of the next, $N+\frac12$th, PN contribution by a $u$-dependent factor $(1-6u)^{-\alpha_{N}}$ which resums the
missing higher-order PN contributions. In the cases considered  in Refs. \cite{Bini:2014zxa,Bini:2016qtx} (which 
dealt with singularities at the lightring), one had some a priori estimates of the value of the exponent $\alpha_{N}$ entering the latter factor. In the cases considered here of singularities at the LSO, we do not have such a priori estimates,
and we shall choose the values of the exponent $\alpha_{N}$ so as to increase the agreement with the numerical SF data
to be discussed next.

Ref. \cite{Akcay:2016dku} has computed numerical values for $\Delta \psi(u_p,e)$ for selected values of $e=[0.05+0.25k]_{k=0\ldots8}$ and $u_p=[(10+5k)^{-1}]_{k=0\ldots18}$. Then Ref. \cite{Kavanagh:2017wot}
extracted (by a fitting procedure) from the latter numerical data, secondary numerical data for the function
$\Delta \psi^{(2)}(u_p)$ parametrizing the $O(e^2)$ contribution to $\Delta \psi(u_p,e)$. 
In Ref. \cite{Kavanagh:2017wot} the latter numerically-derived values of $\Delta \psi^{(2)}(u_p)$
(corresponding to a discrete sample of values of the $u_p$'s)
were denoted $m^{\rm num}(u_p)$, and they were completed by an estimate of a
corresponding numerical (fitting) error, denoted $\sigma^{\rm num}_m(u_p)$. In Ref. \cite{Kavanagh:2017wot}
we had compared the list of numerical data $m^{\rm num}(u_p) \pm \sigma^{\rm num}_m(u_p)$
to the 6PN-accurate theoretical expression for $\Delta \psi^{(2)}(u_p)$  that we had derived there.
Here, we shall investigate to what extent the improved (9PN-accurate) theoretical expression for
$\Delta \psi^{(2)}(u_p)$ derived above improves the agreement between numerical data
and theoretical values. Such a comparative study must crucially take into account both the numerical
error $\sigma^{\rm num}_m(u_p)$ and the relevant theoretical error, as estimated by using the
general formula \eqref{sigmath}. More precisely, when dealing with the
6PN-accurate result of Ref. \cite{Kavanagh:2017wot}, as we (now) know the value of the numerical
coefficient $C_{6.5}^{\psi^{(2)}\!, \,\rm LSO}$, namely 
$C_{6.5}^{\psi^{(2)}\!, \,\rm LSO} \approx - 0.185439$, we shall use its absolute value in defining 
 $\sigma_{6 \, \rm PN}^{\rm th}(\Delta \psi^{(2)}(u))$. In addition, we found that including an extra factor
$ (1-6u)^{-\alpha_{6}}$, with the exponent $\alpha_6=1$, improved the consistency with the numerical data.
In other words, we use $\sigma_{6 \, \rm PN}^{\rm th}(\Delta \psi^{(2)}(u))= 0.185439 (6u)^{6.5}/(1-6u)$.
On the other hand, for the a priori  estimate of the theoretical error on our new 9PN-accurate expression for
$\Delta \psi^{(2)}(u)$, as we do not know the LSO value of the $9.5$PN contribution, we simply use 
as overall numerical coefficient a coefficient equal to 1 (as suggested by the last values in the
second column of Table \ref{tab:1}). In addition, we found that the agreement with numerical data was 
slightly better when using no additional LSO-blowup factor, i.e. we use $\alpha_9=0$. In other words, we simply use
 $\sigma_{9 \, \rm PN}^{\rm th}(\Delta \psi^{(2)}(u)) = (6u)^{9.5}$.
 
The important thing is then to compare the two different numerical-minus-theoretical discrepancies,
say 
\begin{eqnarray} \label{deltanum-th}
\delta^{6 \, \rm PN}(u_p) &\equiv& m^{\rm num}(u_p) - \Delta \psi^{(2) \, 6\rm PN }(u_p) \,, \nonumber\\
\delta^{9 \, \rm PN}(u_p) &\equiv& m^{\rm num}(u_p) - \Delta \psi^{(2) \, 9\rm PN }(u_p) \,,
\end{eqnarray}
to a measure of the total error, combining both the numerical one and the (corresponding)  theoretical one.
As is standard, we define the two total errors corresponding to the two relevant cases (6PN vs 9PN
theoretical accuracies) by summing the two separate errors in quadrature, namely
\begin{eqnarray} \label{sigmatot6PN9PN}
\sigma_{6 \, \rm PN}^{\rm tot}(\Delta \psi^{(2)}(u_p)) &\equiv& \sqrt{(\sigma^{\rm num}_m(u_p))^2 + (\sigma_{6 \, \rm PN}^{\rm th}(u_p))^2}  \,, \nonumber\\ 
\sigma_{9 \, \rm PN}^{\rm tot}(\Delta \psi^{(2)}(u_p)) &\equiv& \sqrt{(\sigma^{\rm num}_m(u_p))^2 + (\sigma_{9 \, \rm PN}^{\rm th}(u_p))^2} \,.\nonumber\\
\end{eqnarray}

\begin{figure}
\includegraphics[scale=0.4]{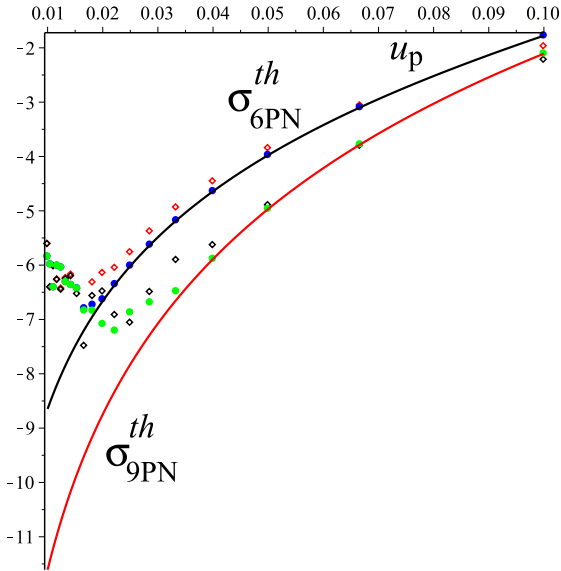}
\caption{\label{FIG1} The data points associated with  $\delta^{6 \, \rm PN}(u_p)$ (diamonds,  red online), $\delta^{9 \, \rm PN}(u_p)$ (diamonds, black online), $\sigma_{6 \, \rm PN}^{\rm tot}$ (solid circles, blue online), $\sigma_{9 \, \rm PN}^{\rm tot}$ (solid circles, green online) are plotted on a semi-logarithmic scale as functions of $u_p$. The two solid curves correspond to (the $\log_{10}$ of) $\sigma^{\rm th}_{\rm 6PN}=0.185439(6u_p)^{6.5}/(1-6u_p)$ and $\sigma^{\rm th}_{\rm 9PN}=(6u_p)^{9.5}$.}
\end{figure}

In Table \ref{tab:2} we present 
the values of the two different numerical-minus-theoretical discrepancies \eqref{deltanum-th}, together with the two corresponding total errors \eqref{sigmatot6PN9PN}.
The corresponding (discrete) data points are plotted (on a semi-logarithmic scale, and using absolute values) in Fig. \ref{FIG1}. In the latter figure, we have also indicated the two continuous curves representing the (base-10 logarithms of the) two theoretical errors \eqref{sigmath}, for $N=6$ and $N=9$. 
Note that each (absolute) value of  $\delta^{N \, \rm PN}(u_p)$
is quite close to the corresponding total error $\sigma_{N \, \rm PN}^{\rm tot}$. More precisely, for $u_p \leq 0.0154$
(i.e. $p\geq 65$) the four different values $\delta^{6 \, \rm PN}(u_p)$, $\delta^{9 \, \rm PN}(u_p)$, $\sigma_{6 \, \rm PN}^{\rm tot}$, $\sigma_{9 \, \rm PN}^{\rm tot}$, are all close to each other, because the theoretical estimates are much closer
to each other than the numerical error, and because they are also in agreement with the numerical data (within the
numerical error). On the other hand, for $u_p > 0.0154$, the data points corresponding to
each separate PN accuracy (6PN vs 9PN) are still close to each other (showing the consistency,
modulo the total error, of each theoretical estimate with the numerical data),
but there is now a notable vertical distance between 
$(|\delta^{6 \, \rm PN}(u_p)|, \sigma_{6 \, \rm PN}^{\rm tot})$, on one side,
and, $(|\delta^{9 \, \rm PN}(u_p)|, \sigma_{9 \, \rm PN}^{\rm tot})$, on the other side.
For this part of the plot, the total error is dominated by the corresponding theoretical one, and we see
that the improved theoretical accuracy is effective in bringing an improved agreement with the
numerical data. This brings a direct numerical confirmation of our new theoretical results.


\begin{table*}
  \caption{
\label{tab:2} 
We compare the two theoretical values $m_i^{\rm thy} \equiv \Delta \psi^{(2) \rm PN}(p_i)$ at 6PN (Ref. \cite{Kavanagh:2017wot}) and 9PN (this work) and the corresponding numerical-minus-theoretical discrepancies $\delta_i$ defined in Eq. \eqref{deltanum-th}.
The estimates of the  corresponding uncertainties in their values are indicated in parenthesis.
The second column shows the numerical estimates $m_i^{\rm num}$ obtained in Ref. \cite{Kavanagh:2017wot} by least-squares fitting the numerical data for $\Delta \psi(p,e)$ given in Ref. \cite{Akcay:2016dku}. 
}
  \begin{center}
    \begin{ruledtabular}
      \begin{tabular}{lccccc}
$p$&   $m^{\rm num}(\sigma^{\rm num}_m)|_{\rm Ref. [25]}$& $m^{\rm thy}(\sigma^{\rm thy}_{m})|_{6PN}$& $\delta^{\rm 6PN}(\sigma^{\rm tot}_{\rm 6PN})$&  $m^{\rm thy}(\sigma^{\rm thy}_{m})|_{9PN}$ & $\delta^{\rm 9PN}(\sigma^{\rm tot}_{\rm 9PN})$\\
\hline
10&      2.83892(11)$\times 10^{-2}$    &     3.9(1.7)$\times 10^{-2}$     & -0.11(17)$\times 10^{-1}$   &  3.44(78)$\times 10^{-2}$         &   - 0.060(78)$\times 10^{-1}$ \cr          
15&      9.12787(61)$\times 10^{-3}$    &     9.99(80)$\times 10^{-3}$     & -8.6(8.0)$\times 10^{-4}$   &  9.28(17)$\times 10^{-3}$        &  -0.16(16)$\times 10^{-3}$ \cr            
20&      4.40237(32)$\times 10^{-3}$    &     4.54(11)$\times 10^{-3}$     & -1.4(1.1)$\times 10^{-4}$   &  4.415(11)$\times 10^{-3}$       &   -0.13(11)$\times 10^{-4}$ \cr          
25&     2.561664(35)$\times 10^{-3}$    &     2.596(23)$\times 10^{-3}$    & -3.5(2.3)$\times 10^{-5}$   &  2.5640(13)$\times 10^{-3}$      &    -0.23(13)$\times 10^{-5}$ \cr          
30&      1.66508(23)$\times 10^{-3}$    &     1.6765(66)$\times 10^{-3}$    & -1.14(66)$\times 10^{-5}$   &  1.66632(23)$\times 10^{-3}$     &   -0.123(33)$\times 10^{-5}$ \cr           
35&      1.16553(20)$\times 10^{-3}$    &     1.1697(24)$\times 10^{-3}$    & -4.1(2.4)$\times 10^{-6}$   &  1.165845(53)$\times 10^{-3}$    &   -3.2(2.1)$\times 10^{-7}$ \cr           
40&      8.5943(13)$\times 10^{-4}$     &     8.6114(96)$\times 10^{-4}$    & -1.71(97)$\times 10^{-6}$   & 8.59511(15)$\times 10^{-4}$      &    -0.86(13)$\times 10^{-7}$\cr          
45&      6.58803(62)$\times 10^{-4}$    &     6.5969(44)$\times 10^{-4}$    & -8.8(4.4)$\times 10^{-7}$   & 6.589218(49)$\times 10^{-4}$     &    -1.20(62)$\times 10^{-7}$ \cr           
50&      5.20356(82)$\times 10^{-4}$    &     5.2107(22)$\times 10^{-4}$    & -7.1(2.3)$\times 10^{-7}$   & 5.206806(18)$\times 10^{-4}$     &  -3.25(82)$\times 10^{-7}$ \cr           
55&      4.2124(14)$\times 10^{-4}$     &     4.2171(12)$\times 10^{-4}$   & -4.8(1.8)$\times 10^{-7}$   & 4.2150318(72)$\times 10^{-4}$    &    -2.7(1.4)$\times 10^{-7}$ \cr          
60&      3.4798(15)$\times 10^{-4}$     &     3.48128(65)$\times 10^{-4}$   & -1.5(1.6)$\times 10^{-7}$   & 3.4800907(32)$\times 10^{-4}$    &   -0.3(1.5)$\times 10^{-7}$ \cr          
65&      2.9178(37)$\times 10^{-4}$     &     2.92146(38)$\times 10^{-4}$   & -3.6(3.7)$\times 10^{-7}$   & 2.9207548(15)$\times 10^{-4}$    &     -2.9(3.7)$\times 10^{-7}$ \cr          
70&      2.4793(43)$\times 10^{-4}$     &     2.48589(24)$\times 10^{-4}$   & -6.6(4.3)$\times 10^{-7}$   &  2.48544960(73)$\times 10^{-4}$  &    -6.2(4.3)$\times 10^{-7}$ \cr          
75&      2.1347(49)$\times 10^{-4}$     &     2.14046(15)$\times 10^{-4}$   & -5.7(4.9)$\times 10^{-7}$   &  2.14017950(38)$\times 10^{-4}$  &      -5.4(4.9)$\times 10^{-7}$ \cr          
80&      1.8583(90)$\times 10^{-4}$     &    1.861997(98)$\times 10^{-4}$   & -3.7(9.0)$\times 10^{-7}$   &  1.86181379(21)$\times 10^{-4}$  &      -3.5(9.0)$\times 10^{-7}$ \cr          
85&      1.6289(97)$\times 10^{-4}$     &    1.634304(66)$\times 10^{-4}$   & -5.4(9.7)$\times 10^{-7}$   & 1.63417981(12)$\times 10^{-4}$   &   -5.3(9.7)$\times 10^{-7}$ \cr           
90&      1.4552(39)$\times 10^{-4}$     &    1.445785(45)$\times 10^{-4}$   & +9.5(3.9)$\times 10^{-7}$   & 1.445699417(67)$\times 10^{-4}$  &      +9.5(3.9)$\times 10^{-7}$ \cr         
95&      1.292(10)$\times 10^{-4}$      &    1.287970(32)$\times 10^{-4}$   & +0.4(1.0)$\times 10^{-6}$   &  1.287910310(40)$\times 10^{-4}$ &   +0.4(1.0)$\times 10^{-6}$ \cr          
100&     1.130(14)$\times 10^{-4}$      &    1.154556(23)$\times 10^{-4}$   & -2.4(1.4)$\times 10^{-6}$   & 1.154513310(25)$\times 10^{-4}$  &   -2.4(1.4)$\times 10^{-6}$ \cr          
\end{tabular}
\end{ruledtabular}
\end{center}
\end{table*}


The analysis above has provided us with an estimate of the theoretical error on our new, 9PN-accurate
result for $\Delta \psi^{(2)}(u_p)$, namely the function 
$\sigma_{9 \, \rm PN}^{\rm th}(\Delta \psi^{(2)}(u)) = (6u)^{9.5}$. This theoretical error gets 
large as $u$ approaches $\frac16$.
More precisely, one finds that the fractional theoretical error
\beq
\frac{ \sigma_{9 \, \rm PN}^{\rm th}(\Delta \psi^{(2)}(u))}{\Delta \psi^{(2) \, 9 \rm PN}(u)}
\eeq
increases monotonically with $u$, to reach $ 12.4\%$ when $u=0.09$, $ 22.7\%$ when $u=0.1$,
and $ 36.7\%$ when $u=0.11$. This illustrates again the poor convergence of PN approximants. 
Here, the situation is worse than usual because, as we argued above, the function $\Delta \psi^{(2)}(u)$
probably has a singularity at (or near) $u=u_{\rm LSO}= \frac16 = 0.1666\ldots$. Even our 9PN-accurate expansion
becomes useless above $u \simeq 0.1$. 

\begin{figure*}
\includegraphics[scale=0.33]{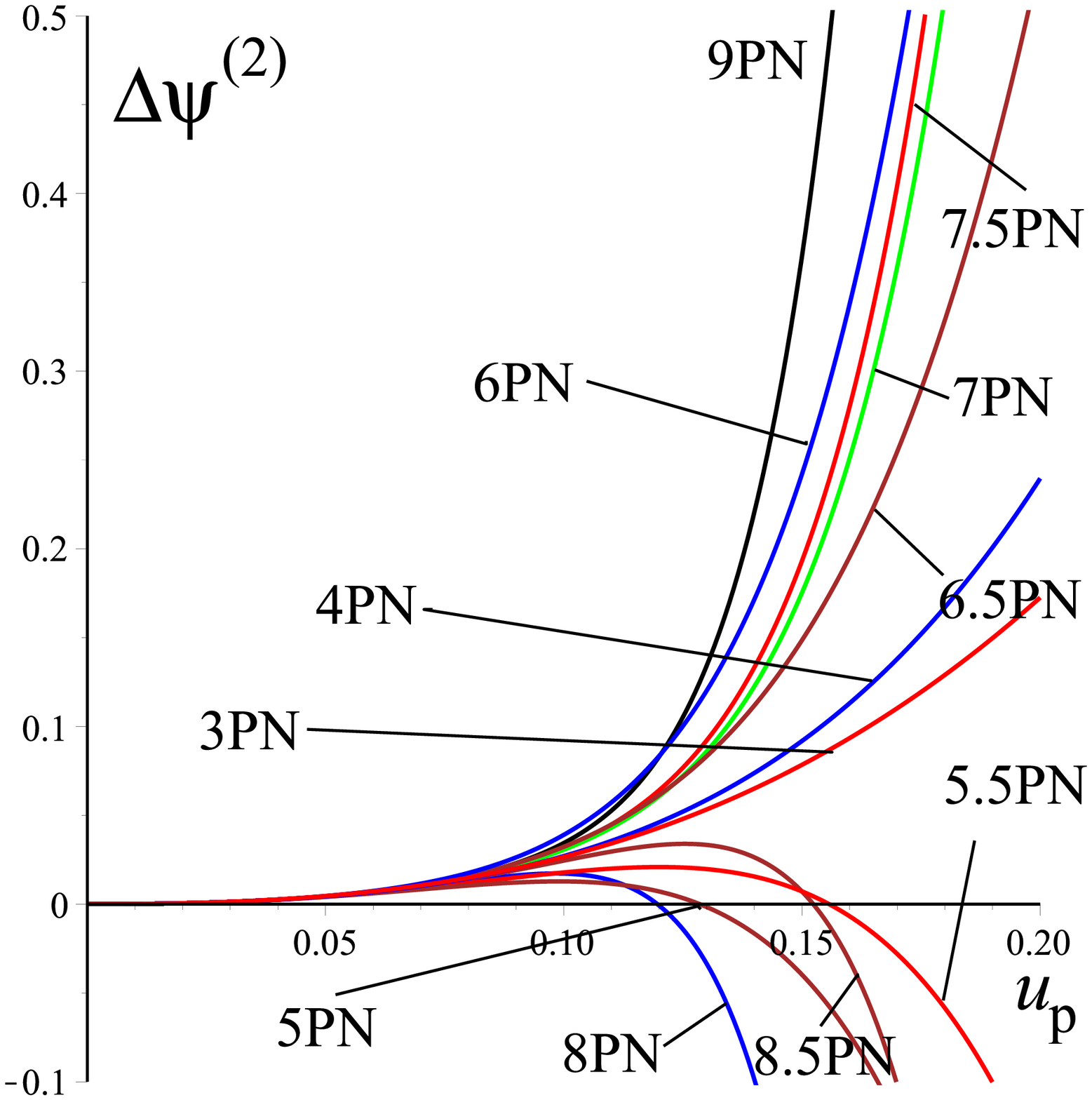}\qquad\qquad
\includegraphics[scale=0.33]{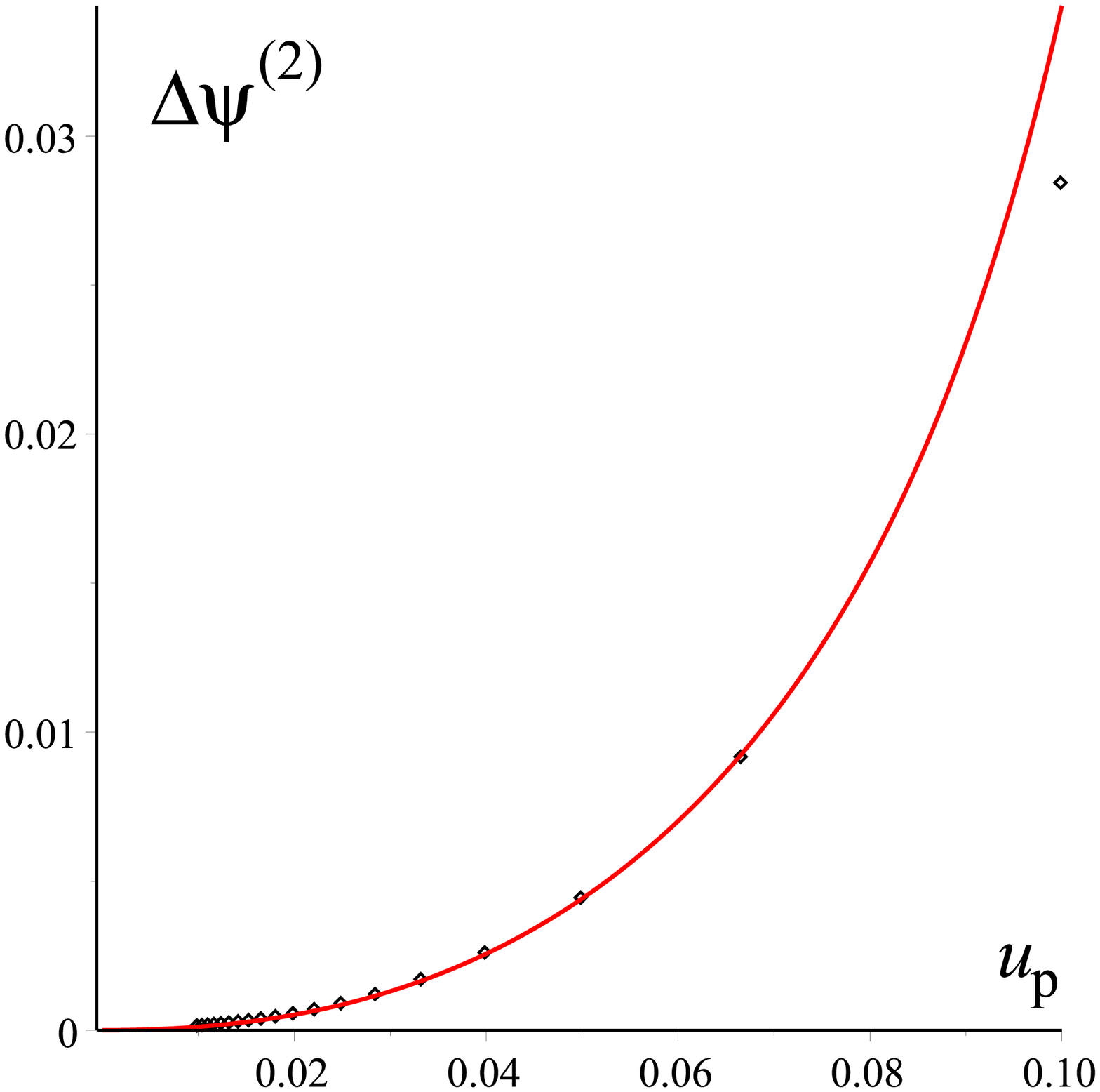}
\caption{\label{fig:1} (Left panel) The various PN-approximants to $\Delta \psi^{(2)}(u_p)$, Eq. \eqref{psi2NPN} with $ 3 \leq N \leq 9$, are plotted as functions of the inverse-semilatus rectum $u_p$.
(Right panel) Our 9PN result for $\Delta \psi^{(2)}(u_p)$ superposed to the corresponding data extracted 
(in Ref. \cite{Kavanagh:2017wot}) from Ref.  \cite{Akcay:2016dku}.
}
\end{figure*}

In Fig. \ref{fig:1}, left panel, we plot the sequence of $N$-PN approximants
to $\Delta \psi^{(2)}(u)$,
as defined in Eq. \eqref{psi2NPN}, for $N \geq 3$. In the right panel, instead we compare our 9PN result for $\Delta \psi^{(2)}(u)$ to the numerical data extracted from Ref. \cite{Akcay:2016dku}.  

As there are no numerical data for $u_p > 0.1$, and as the current
theoretical estimates get (as explained above) completely unreliable for $u_p > 0.1$, we see that
we have no firm knowledge of the behavior of $\Delta \psi^{(2)}(u)$ for $u_p \gtrsim 0.1$. The enormous
spread among the various PN approximants cannot reliably tell us whether $\Delta \psi^{(2)}(u)$ goes
to $ +\infty$ or $ -\infty$ (or has a milder behavior) as $u$ approaches $\frac16$.


Finally, let us discuss the convergence properties of the PN expansion of the 1SF
contributions to the EOB gyrogravitomagnetic ratio $g_{S*}(u, p_r, p_\phi,\nu)$.
We recall that, according to Eq. \eqref{nuexpgS}, the SF expansion (i.e. the expansion in powers
of $\nu$) of $g_{S*}(u,p_r,p_\phi,\nu)$ is decomposed into the zeroth contribution \eqref{gSs0}
(expressed as a specific function of $u, p_r$ and $p_\phi$), and into a 1SF contribution \eqref{gSs1SF}
which is expanded in powers of $p_r^2$. Here, we expect different radii of convergence for the
PN expansions of the various contributions $g_{S*}^{\rm 1SF0}(u), g_{S*}^{\rm 1SF2}(u), \ldots$
to the 1SF term $g_{S*}^{(\nu^1)}(u,p_r)$. Indeed, when taking from the start the limit $p_r \to 0$,
i.e., when considering the sequence of circular orbits, the only place where a singularity  can appear is
at the LR, i.e., for $u= u_{\rm LR}=\frac13$. We then expect that the successive PN terms in the
PN expansion of the function $g_{S*}^{\rm 1SF0}(u)$ will be of order 1 at  $u= u_{\rm LR}=\frac13$.
The first column of Table \ref{tableCnpsilso} lists the separate, successive PN contributions to $g_{S*}^{\rm 1SF0}(u)$,
evaluated at $u=\frac13$, and we see that, indeed, they are roughly all of order unity.

On the other hand, because of the specific gauge choice we made of writing (for definiteness) the 
coefficient $g_{S*}^{\rm 1SF2}(u)$ of the $O(p_r^2)$ contribution to $g_{S*}^{(\nu^1)}(u,p_r)$
as a function only of $u$ (rather than of both $u$ and $p_\phi$), it is to be expected that the
function $g_{S*}^{\rm 1SF2}(u)$ will inherit from its ``source"  function $\Delta \psi^{(2)}(u_p)$
the presence of a singularity at the LSO, i.e. at $u= u_{\rm LSO}=\frac16$. Indeed, even if we
assume the existence of some unknown EOB function $g_{S*}^{(\nu^1)}(u,p_r, p_\phi)$
that would hopefully only be singular at the LR (but be regular at the LSO), one needs
to apply a gauge transformation to gauge-fix $g_{S*}^{(\nu^1)}(u,p_r, p_\phi)$
into the form $g_{S*}^{(\nu^1)}(u,p_r)$, and the determination
of this transformation must involve the comparison of gauge-invariant functions of two variables,
i.e., functions of the two frequencies $\Omega_r$ and $\Omega_\phi$. For the reasons explained
above, the latter comparison will then introduce an extra singularity at $u= u_{\rm LSO}=\frac16$.
In the second column of Table \ref{tableCnpsilso} we list the separate, successive PN contributions to $g_{S*}^{\rm 1SF2}(u)$,
evaluated at $u=\frac16$, and we see that, indeed, they stay roughly all of order unity (possibly except for
the last one, which is largish). By contrast, when evaluating the successive PN contributions to $g_{S*}^{\rm 1SF2}(u)$,
evaluated at $u=\frac13$, we found that they became increasingly large as the PN order increases
(for instance the $O(u^6)$ contribution is equal to $-75.38253$ at $u=\frac13$,
while the $O(u^7)$ one is equal to $928.63276$).


\begin{table}[t]
  \caption{
\label{tableCnpsilso} 
Numerical values of the successive $O_{\ln}(u^n)$ contributions to
both $g_{S*}^{\rm 1SF0}(u_p)$  (evaluated at $u=u_{\rm LR}=1/3$),
and $g_{S*}^{\rm 1SF2}(u_p)$ (evaluated at $u=u_{\rm LSO}=1/6$).
}
  \begin{center}
    \begin{ruledtabular}
      \begin{tabular}{|l|c|c|}
\hline 
$PN$ & $g_{S*}^{\rm 1SF0}(1/3)$ & $g_{S*}^{\rm 1SF2}(1/6)$ \cr
\hline 
$\eta^0$      & -                     & -2.25 \cr
$\eta^2$      & -0.25                    & -0.375 \cr 
$\eta^4$      & -1.083333   &  -0.622396  \cr 
$\eta^6$      & -1.072353   &  +0.487256$\times 10^{-2}$  \cr 
$\eta^8$      & -0.421860   &  +0.820994 \cr 
$\eta^9$      & -   &  -0.224519  \cr
$\eta^{10}$   & +0.748491   &  -0.919538  \cr 
$\eta^{11}$   & -0.442184   &  +0.350432  \cr 
$\eta^{12}$  &  -2.099104   &  -0.638599  \cr 
$\eta^{13}$   & +0.775711   &  +0.569630$\times 10^{-1}$  \cr 
$\eta^{14}$  &  -1.520959   &  +6.602989  \cr 
\end{tabular}
\end{ruledtabular}
\end{center}
\end{table}

\begin{figure}
\includegraphics[scale=0.33]{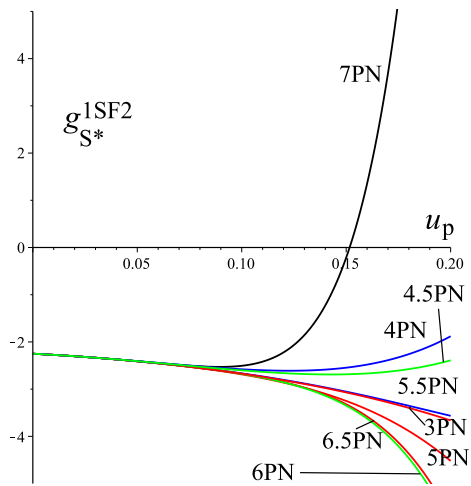}
\caption{\label{fig:2} The various PN-approximants to $g_{S*}^{\rm 1SF2}(u_p)$, Eq. \eqref{gsstar1SF2}, are plotted as functions of the inverse-semilatus rectum $u_p$.
}
\end{figure}

Using the same reasoning we employed above to estimate  the theoretical error on 
the truncated PN expansions of $\Delta \psi^{(2)}(u)$, we then expect that a reasonable estimate of the
theoretical error on the current 7PN-accurate PN expansion of $g_{S*}^{\rm 1SF2}(u)$,
Eq. \eqref{gsstar1SF2}, i.e., an estimate of the error term $O_{\ln{}}(u^{7.5})$  in the latter equation,
is roughly given by
\beq
\sigma^{\rm th}_{7 \rm PN}(g_{S*}^{\rm 1SF2}(u))= (6 u)^{7.5}/(1-6u).
\eeq
The latter error goes to infinity when $u$ approaches $\frac16=0.1666\ldots$, and becomes
already unacceptably large around $u \simeq 0.13$. Indeed, the (absolute value of the) fractional error
$\sigma^{\rm th}(g_{S*}^{\rm 1SF2}(u))/ g_{S* , 7 \rm PN}^{\rm 1SF2}(u)$ increases with $u$
and is found to be equal to $13.68 \%$ when $u=0.12$, and $38.00\%$ when $u=0.13$.
In other words, even the  much improved 7PN-accurate expansion of $g_{S*}^{\rm 1SF2}(u)$
derived in the present work becomes totally unreliable for $u > 0.12$, so that we do not have
any solid knowledge of the strong-field behavior of $g_{S*}^{\rm 1SF2}(u)$. In absence of
direct numerical data on $g_{S*}^{\rm 1SF2}(u)$ we have no firm knowledge of the behavior
of this function beyond $u=0.12$, and, in particular, of its probable singularity structure at $u=\frac16$.
One would need an analytical knowledge of the latter singularity structure in order to concoct
a more regular version, say $g_{S*}^{\rm 1SF2}(u, p_\phi)$, involving some dependence on $p_\phi$.
The enormous strong-field spread among the various PN approximants to $g_{S*}^{\rm 1SF2}(u)$
is illustrated in Fig. \ref{fig:2}.


\section{Concluding remarks}

Improving upon recent results by Kavanagh et al. \cite{Kavanagh:2017wot}, we have analytically computed,
 through the ninth post-Newtonian (PN) order, the $O(e^2)$ contribution to the
(first-order) gravitational self-force (SF) correction to the spin-orbit precession of a spinning compact body along a slightly
 eccentric orbit around a Schwarzschild black hole (see Eq. \eqref{result}). We have then translated this information
 into its corresponding Effective-One-Body (EOB) counterpart, thereby determining through the (fractional)
 seventh PN order the $O(p_r^2)$ self-force contribution to 
the EOB gyrogravitomagnetic ratio $g_{S*}$ (see Eq. \eqref{gsstar1SF2}).
We have shown the compatibility between our improved analytical knowledge of $\Delta \psi^{(2)}$ and numerical
SF data extracted in Ref. \cite{Kavanagh:2017wot} from the numerical results of Ref. \cite{Akcay:2016dku}
(see Fig. \ref{FIG1} and Table \ref{tab:2}). We have studied the convergence of the PN expansions of both $\Delta \psi^{(2)}(u)$ 
and $g_{S*}^{1SF2}(u)$ and emphasized that their convergence is much worse than that of the
usual, circular-orbit related dynamical quantities. Indeed, the existence, in the unperturbed
background spacetime, of a Last Stable (circular) Orbit (LSO) at $r=6GM/c^2$ implies the presence of
a singularity at $u=\frac16$ in the (exact) functions $\Delta \psi^{(2)}(u)$ and $g_{S*}^{1SF2}(u)$, and, this singularity
then entails that the radius of convergence of the PN expansions of $\Delta \psi^{(2)}(u)$ and $g_{S*}^{1SF2}(u)$
is only equal to $u_{\rm LSO}=\frac16$. This radius of convergence is twice smaller than that of the
usual, circular-orbit related dynamical potentials (such as the SF contribution $a_{1SF}(u)$ to the main
EOB radial potential). The resulting bad convergence of the sequence of PN approximants has been
illustrated in Figs. \ref{fig:1} and \ref{fig:2}. If one wants to overcome this problem, one would need to study
the precise analytical structure of the LSO singularity of the functions $\Delta \psi^{(2)}(u)$ and $g_{S*}^{1SF2}(u)$.
We leave this task to future work, as well as the technically challenging task of further extending our results
in the following directions: computing higher PN orders, including higher order contributions in eccentricity,
and taking into account  the spin of the central black hole.

\appendix

\section{Definition of the spin-precession invariant}

\subsection{Gyroscope precession in the background spacetime}

The tangent 4-velocity $\bar u$ ($\bar u\cdot\bar u=-1$) to an unpertubed eccentric geodesic orbit on the equatorial plane of the background Schwarzschild spacetime is given by 
\beq
\bar u=\bar u^\alpha\partial_\alpha=\frac{\bar E}{f}\partial_t + \dot r \partial_r +\frac{\bar L}{r^2}\partial_\phi\,, 
\eeq
where $\dot r \equiv \bar u^r$ is such that
\beq
\dot r^2=\left(\frac{dr}{d\bar\tau}\right)^2
=\bar E^2-f\left(1+\frac{\bar L^2}{r^2}\right)\,.
\eeq
The orbit can be parametrized either by the proper time $\bar \tau$ or by the relativistic anomaly $\chi\in[0,2\pi]$, such that 
\beq
r=\frac{m_2 p}{1+e\cos \chi}\,.
\eeq
They are related by 
\beq
\label{dtaudchi}
\frac{d\bar\tau}{d\chi}=\frac{m_2p^{3/2}}{(1+e\cos\chi)^2}\left[\frac{p-3-e^2}{p-6-2e\cos\chi}\right]^{1/2}\,.
\eeq
The (dimensionless) background orbital parameters, semi-latus rectum $p$ and eccentricity $e$, are defined by writing the minimum (pericenter,
$r_{\rm peri}$) and maximum (apocenter, $r_{\rm apo}$) values of the (areal) radial coordinate along the orbit as 
\beq
r_{\rm peri}=\frac{m_2 p}{1+e}\,,\qquad r_{\rm apo}=\frac{m_2 p}{1-e}\,.
\eeq
They are in correspondence with the conserved (dimensionless) energy $\bar E=-\bar u_t$ and angular momentum $\bar L= \bar u_\phi$ per unit mass of the particle, via
\beq
\label{energy}
\bar E^2= \frac{ (p-2)^2-4e^2  }{p(p-3-e^2)}\,,\qquad 
\bar L^2=\frac{p^2 }{p-3-e^2}\,.
\eeq
The reciprocal of $p$, $u_p  \equiv p^{-1}$, is  a useful argument, which serves also
as PN expansion parameter.
 
Eq. (\ref{dtaudchi}) can be used to solve the equations for $t$ and $\phi$ as functions of $\chi$, which are 
then solvable in terms of elliptic functions.  
As  is well known, eccentric orbits are characterized by two fundamental frequencies,  $\bar\Omega_{r}=2\pi/\bar T_{r}$ and $\bar\Omega_{\phi}=\bar\Phi/\bar T_{r}$, where $\bar\Phi=\oint d\phi = \oint d\chi d\phi/d\chi $ is the angular advance during one radial period, $\bar T_{r}=\oint dt= \oint d\chi dt/d\chi  $. 
To second order in $e$ we find
\begin{eqnarray}
m_2 \bar\Omega_{r} &=& u_p^{3/2}(1-6u_p)^{1/2}\left[ \right.\nonumber\\
&& \left.1-\frac34 \frac{2-32u_p+165u_p^2-266u_p^3}{ (1-2u_p) (1-6u_p)^2}\, e^2
+O(e^4)\right]\,,\nonumber\\
m_2 \bar\Omega_{\phi} &=& u_p^{3/2}\left[1-\frac32 \frac{1-10u_p+22u_p^2}{(1-2u_p)(1-6u_p)}\, e^2\right.\nonumber\\
&& \left.
+O(e^4)\right]\,.
\end{eqnarray}

The gyroscope precession is defined with respect to the Marck-type frame \cite{Marck} adapted to $\bar u$, completed by the spatial triad
\begin{eqnarray}
\bar e_1&=& 
\frac{1}{\sqrt{1+\bar L^2/r^2}}\left[\frac{\dot r }{f}\partial_t + \bar E  \partial_r\right]\,,\nonumber\\
\bar e_2&=& \frac{1}{r}\partial_\theta\,, \nonumber\\
\bar e_3
&=& \frac{1}{\sqrt{1+\bar L^2/r^2}}\left[\frac{\bar L}{r}\bar u +\frac{1}{r} \partial_\phi\right] \,,
\end{eqnarray}
whose transport properties are
\beq
\nabla_{\bar u} \bar e_1 =\bar \omega \bar e_3\,,\qquad 
\nabla_{\bar u} \bar e_3 =-\bar \omega \bar e_1\,,
\eeq
with
\beq
\bar \omega=\frac{\bar E\bar L}{r^2+\bar L^2}\,. 
\eeq
The precession angle of a test gyroscope dragged along $\bar u$ is then given by
\beq
\bar \psi=1-\frac{\bar \Psi}{\bar \Phi}\,,
\eeq
where
\beq
\bar \Psi=\int_0^{\bar{\mathcal T}} \bar \omega d\bar \tau
=\int_0^{2\pi} \bar \omega \frac{d\bar \tau}{d\chi}d\chi\,,
\eeq
which finally yields
\begin{eqnarray}
\bar \psi&=&1-\sqrt{1-3u_p}\nonumber\\
&&+\frac{3}{2}\frac{(1-4u_p)u_p^2}{(1-2u_p)(1-6u_p)\sqrt{1-3u_p}}\,e^2\nonumber\\
&&+O(e^4)
\,.
\end{eqnarray}

\subsection{Spin precession in the perturbed spacetime}

Bound timelike geodesics in the equatorial plane of the perturbed spacetime have $4$-velocity 
\begin{eqnarray}
u&=&u^\alpha\partial_\alpha=(\bar u^\alpha+\delta u^\alpha)\partial_\alpha\\
&=&\frac1{f}(\bar E+\delta E)\partial_t + (\bar u^r+\delta u^r) \partial_r +\frac1{r^2}(\bar L+\delta L)\partial_\phi\,,\nonumber
\end{eqnarray}
with $\delta u^\alpha =O(h)$.
Here, $\delta u^r$ follows from the normalization condition of $u$ with respect to the perturbed metric, which reads
\beq
\label{delta_ur}
\bar u^r \delta u^r = \bar E \delta E -\frac{\bar L}{r^2}f \delta L -\frac12 f h_{00}\,,
\eeq
where $h_{00}=h_{\alpha\beta}\bar u^\alpha \bar u^\beta$.
Equivalently, one can normalize $u$ with respect to the background metric as in Barack and Sago (BS) \cite{Barack:2011ed}, leading to 
\beq
\label{delta_ur_BS}
\bar u^r \delta u^r_{BS} = \bar E \delta E_{BS} -\frac{\bar L}{r^2}f \delta L_{BS}\,,
\eeq
where
\begin{eqnarray}
\label{relwithBS}
\delta E_{BS}&=&\delta E-\frac12\bar E h_{00}\,, \nonumber\\
\delta u^r_{BS}&=&\delta u^r-\frac12\bar u^r h_{00}\,, \nonumber\\
\delta L_{BS}&=&\delta L-\frac12\bar L h_{00}\,.
\end{eqnarray}
The 4-velocity 1-form turns out to be
\begin{eqnarray}
u^\flat&=&u_\alpha dx^\alpha\nonumber\\
&=&-({\bar E}+\delta E-h_{t\bar u})dt +\frac1{f}(\dot r+\delta u^r +fh_{r\bar u}) dr \nonumber\\
&& +({\bar L}+\delta L+h_{\phi \bar u}) d\phi\,,
\end{eqnarray}
where $h_{\alpha\bar u}=h_{\alpha\beta}\bar u^\beta$, and where the further equatorial plane condition $\delta u_\theta=0$ (implying $h_{\theta\bar u}=0$) has been assumed.

The geodesic equations 
\beq
\frac{du_\alpha}{d\tau}-\frac12 (\bar g_{\lambda\mu,\alpha}+h_{\lambda\mu,\alpha})u^\lambda u^\mu=0\,,
\eeq  
determine the evolution of $\delta u_t$ and $\delta u_\phi$, or equivalently of the perturbations in
energy $\delta E$ and angular momentum $\delta L$ by
\begin{eqnarray}
\label{eqdeltaEL}
\frac{d}{d\tau}\delta E &=&\frac12\bar E\frac{dh_{00}}{d\tau}-F_t\,,\nonumber\\  
\frac{d}{d\tau}\delta L &=&\frac12\bar L\frac{dh_{00}}{d\tau}+F_\phi\,,
\end{eqnarray}
where the functions $F_t$ and $F_\phi$ are the covariant $t$ and $\phi$ components of the self force
\beq
F^\mu=-\frac12(\bar g^{\mu\nu}+\bar u^\mu\bar u^\nu)\bar u^\lambda\bar u^\rho(2h_{\nu\lambda;\rho}-h_{\lambda\rho;\nu})\,.
\eeq
Here we are interested in conservative effects only, i.e., we assume that $F^\alpha=F^\alpha_{\rm cons}$ results in a periodic function of $\chi$.
Eqs. \eqref{eqdeltaEL} can then be formally integrated as
\begin{eqnarray}
\delta E_{BS}(\chi) &=& -\int_0^\chi F_t^{\rm cons}(\chi) \frac{d\tau}{d\chi}d\chi+\delta E_{BS}(0)
\nonumber\\
&\equiv& {\mathcal E}_{BS}(\chi)+\delta E_{BS}(0)
\,,\nonumber\\
\delta L_{BS}(\chi) &=& \int_0^\chi F_\phi^{\rm cons}(\chi) \frac{d\tau}{d\chi}d\chi+\delta L_{BS}(0)
\nonumber\\
&\equiv& {\mathcal L}_{BS}(\chi)+\delta L_{BS}(0)
\,,
\end{eqnarray}
recalling the relations \eqref{relwithBS}. Here,  the conservative self force components are defined by $F_t^{\rm cons}=[F_t(\chi)-F_t(-\chi)]/2$ and $F_\phi^{\rm cons}=[F_\phi(\chi)-F_\phi(-\chi)]/2$.
The integration constants $\delta E_{BS}(0)$ and $\delta L_{BS}(0)$ are computed as indicated in Ref. \cite{Barack:2011ed}, and turn out to be
\begin{eqnarray}
\label{deltaELperiBS}
\delta E_{BS}(0)&=& \frac{(1+e)^2 (p-2-2e)}{4e(p-3-e^2)}\times \nonumber\\
&& [(1-e)^2(p-2+2e)B {\mathcal L}_{BS}(\pi) -{\mathcal E}_{BS}(\pi)]
\,,\nonumber\\
\delta L_{BS}(0)&=&  \frac{1}{4e(p-3-e^2)B}\times \nonumber\\
&& \left[(1-e)^2(p-2+2e) B{\mathcal L}_{BS}(\pi) -{\mathcal E}_{BS}(\pi)\right]
\,,\nonumber\\
\end{eqnarray}
with
\beq
B=\frac{1}{m_2^2p^3}\, \frac{\bar L}{\bar E}=\frac{1}{m_2p^{3/2}[(p-2)^2-4e^2]^{1/2}}\,.
\eeq

The spin precession has been calculated by Akcay et al. \cite{Akcay:2016dku} with respect to a suitably defined perturbed Marck-type frame $\{u,e_a\}$ adapted to $u$, with $e_a^\alpha=\bar e_a^\alpha +\delta e_a^\alpha$.
The first-order SF correction to the spin precession invariant turns out to be given by
\beq
\Delta \psi=-\frac{\Delta \Psi}{\bar \Phi}\,,
\eeq
where
\beq
\Delta \Psi=\delta \Psi-\frac{\partial \bar \Psi}{\partial \bar \Omega_r}\delta \Omega_r -\frac{\partial \bar \Psi}{\partial \bar \Omega_\phi}\delta \Omega_\phi\,.
\eeq
The SF corrections to the frequencies are given by
\beq
\delta \Omega_r=-\bar \Omega_r\frac{\delta T}{\bar T}\,,\qquad
\delta \Omega_\phi=-\bar \Omega_\phi\left(-\frac{\delta \Phi}{\bar \Phi}+\frac{\delta T}{\bar T}\right)\,,
\eeq 
where 
\begin{eqnarray}
\label{deltaTePhi}
\delta T&=&\int_0^{2\pi}  \left(\frac{\delta u^t}{\bar u^t}-\frac{\delta u^r}{\bar u^r}\right) \bar u^t \frac{d \bar \tau}{d\chi}  d\chi\nonumber\\
&=&\int_0^{2\pi}  \left(\frac{\delta E}{\bar E}-\frac{\delta u^r}{\bar u^r}\right) \frac{\bar E}{f} \frac{d \bar \tau}{d\chi}  d\chi\,, \nonumber\\
\delta \Phi&=&\int_0^{2\pi}  \left(\frac{\delta u^\phi}{\bar u^\phi}-\frac{\delta u^r}{\bar u^r}\right) \bar u^\phi \frac{d \bar \tau}{d\chi}  d\chi\nonumber\\
&=&\int_0^{2\pi}  \left(\frac{\delta L}{\bar L}-\frac{\delta u^r}{\bar u^r}\right) \frac{\bar L}{r^2} \frac{d \bar \tau}{d\chi}  d\chi\,.
\end{eqnarray}
Finally
\begin{eqnarray}
\delta \Psi&=&\int_0^{2\pi}  \left(\frac{\delta \omega}{\bar \omega}-\frac{\delta u^r}{\bar u^r}\right) \bar \omega \frac{d \bar \tau}{d\chi}  d\chi\\
&\equiv&\int_0^{2\pi} \left[\delta_{h}+\delta_{\partial h}+c_{\delta E_{BS}}\delta E_{BS}+c_{\delta L_{BS}}\delta L_{BS}\right] \frac{d \bar \tau}{d\chi}  d\chi\,,\nonumber
\end{eqnarray}
with
\begin{eqnarray}
c_{\delta E_{BS}}&=&-\frac{\bar L}{r^2}+\frac{(r^2+\bar L^2)(3M-r)\bar L}{r^5(\bar u^r)^2}
\,,\nonumber\\
c_{\delta L_{BS}}&=& \frac{\bar E}{r^2}-\frac{(3 M-r) \bar E\bar L^2 }{ r^5(\bar u^r)^2}
\,,\nonumber\\
\delta_{h}&=& \frac{\bar E \bar L}{2 r^3}\left[(3 M-r)  h_{rr}   
-\frac{  M}{f^2}  h_{tt} -\frac{1}{r}  h_{\phi\phi} \right]\nonumber\\
&&+\left[\frac{  (4 M-r)\bar L}{r f}  h_{tr} +\bar E  h_{r\phi}\right]\frac{\bar u^r}{2r^2} \nonumber\\
&& +\left[-\frac{(4M-r)\bar E^2}{f} +\frac{(5M-2r)\bar L^2}{ r^2} \right.\nonumber\\
&& \left.+ (3M-r)  \right]\frac{ h_{t\phi} }{2 r^3 f}
\,,\nonumber\\
\delta_{\partial h}&=& \left[\left(\frac{h_{t[\phi,t]}}{f^2}   + h_{r[\phi,r]} \right)\bar E \right.\nonumber\\
&& \left.+\left( \frac{h_{\phi[\phi,t]}}{r^2f}  - h_{r[r,t]}     \right)\bar L\right] \frac{\bar u^r}{r}\nonumber\\
&& +\frac{\bar E^2  h_{t[\phi,r]}+(\bar u^r)^2 h_{r[\phi, t]}}{r f}+\frac{\bar L^2}{r^3}h_{\phi[t,r]}\nonumber\\
&&-\left(\frac{h_{\phi [r,\phi]}}{r^2} +\frac{h_{t[r,t]}}{ f}  \right)\frac{\bar E \bar L}{r}
\,.
\end{eqnarray}

\section{Self-force calculation}

In order to obtain the metric perturbation we closely follow the approach of Kavanagh et al. \cite{Kavanagh:2017wot}, who used a radiation gauge and a related Teukolsky formalism.
The set of PN solutions to the Teukolsky radial equation with $s=2$ together with the Mano-Suzuki-Takasugi \cite{Mano:1996vt} solutions for $l=2,\ldots,7$ are used to reconstruct the metric for $l\geq2$. This allows one to compute the $t$ and $\phi$ components of the conservative self-force needed to calculate the perturbed orbit quantities $\delta E$ and $\delta L$, the induced shift of the orbital frequencies $\delta \Omega_r$ and $\delta \Omega_\phi$, the variation $\delta \Psi$ of the accumulated phase of the spin vector, and finally the spin-precession invariant $\delta\psi$.
The contribution of the multipoles $l=0,1$ corresponding to the spacetime perturbations due to the mass and angular momentum of the small body is computed separately.
Finally, the so obtained value of $\delta\psi$ has to be regularized by subtracting out its singular part. 
We refer to Ref. \cite{Kavanagh:2017wot} for a detailed account of all these intermediate steps and provide below only the relevant information about nonradiative multipoles, and the regularization parameter used in our analysis.

\subsection{Low multipoles}

The contribution of the lowest modes $l=0,1$ is obtained by using the solution for the interior and exterior perturbed metric given in Appendix A of Ref. \cite{Hopper:2015icj} by using the Regge-Wheeler-Zerilli approach.
We find
\begin{widetext}
\begin{eqnarray}
	\Delta\psi^+_{l=0}&=&
	\frac{(-1+2u_p)u_p^2(14u_p-3)}{(86u_p^2-39u_p+4)(-1+3u_p)}\nonumber\\
&&
	-\frac{u_p^2(132888u_p^7-273260u_p^6+252318u_p^5-129169u_p^4+38665u_p^3-6710u_p^2+624u_p-24)}{2(-1+6u_p)(-1+3u_p)^2(86u_p^2-39u_p+4)^2(-1+2u_p)}e^2
+O(e^4)	\,,\nonumber \\
	\Delta\psi^-_{l=0}&=& 
	-\frac{(14u_p-3)u_p^3}{(86u_p^2-39u_p+4)(-1+3u_p)}\nonumber\\
&&
	+\frac{(131460u_p^6-247960u_p^5+177821u_p^4-63837u_p^3+12278u_p^2-1210u_p+48)u_p^3}{2(-1+6u_p)(-1+3u_p)^2(86u_p^2-39u_p+4)^2(-1+2u_p)}e^2
+O(e^4)	\,,
\end{eqnarray}
and
\begin{eqnarray}
	\Delta\psi^+_{\ell=1}&=&
	-\frac{u_p(14u_p^3+55u_p^2-33u_p+4)}{(86u_p^2-39u_p+4)(-1+3u_p)}\nonumber\\
&&
	+\frac{(261492u_p^7-583264u_p^6+518173u_p^5-241971u_p^4+65133u_p^3-10243u_p^2+880u_p-32)u_p^2}{(-1+6u_p)(-1+3u_p)^2(86u_p^2-39u_p+4)^2(-1+2u_p)}e^2
+O(e^4)		\,,\nonumber \\
	\Delta\psi^-_{\ell=1}&=&
	\frac{2(-1+2u_p)(28u_p^2-17u_p+2)u_p}{(86u_p^2-39u_p+4)(-1+3u_p)}\nonumber\\
&&
	-\frac{(531552u_p^7-980396u_p^6+772244u_p^5-337047u_p^4+87696u_p^3-13517u_p^2+1136u_p-40)u_p^2}{2(-1+6u_p)(-1+3u_p)^2(86u_p^2-39u_p+4)^2(-1+2u_p)}e^2
+O(e^4)		\,.
\end{eqnarray}

\subsection{Regularization}

To regularize the quantity $\Delta\psi$, it is enough to subtract the large-$l$ limit of its PN expansion, i.e.,
\beq
\Delta\psi=\sum_{\ell=0}^{\infty}\left[\frac{1}{2}\left(\Delta\psi^{l,+}+\Delta\psi^{l,-}\right)-B\right]\,,
\eeq
where the left and right contributions are such that $\Delta\psi^{l,+}=\Delta\psi^{-l-1,-}$ and 
\beq
B(u_p,e)=B_0(u_p)+e^2B_2(u_p)\,,
\eeq
with 
\begin{eqnarray}
B_0(u_p)&=&\frac{21}{16}u_p-\frac{201}{128}u_p^2+\frac{529}{1024}u_p^3+\frac{152197}{16384}u_p^4+\frac{17145445}{262144}u_p^5+\frac{886692225}{2097152}u_p^6+\frac{45206277105}{16777216}u_p^7\nonumber\\
&&
+\frac{9204713714385}{536870912}u_p^8+\frac{1875482334818445}{17179869184}u_p^9
+O(u_p^{10})\,,
\end{eqnarray}
and
\begin{eqnarray}
B_2(u_p)&=&-\frac{435}{512}u_p^2-\frac{1155}{1024}u_p^3-\frac{352849}{65536}u_p^4-\frac{5100243}{131072}u_p^5-\frac{2456459237}{8388608}u_p^6-\frac{36003649389}{16777216}u_p^7\nonumber\\
&&
-\frac{32713771158557}{2147483648}u_p^8-\frac{451723973383879}{4294967296}u_p^9
+O(u_p^{10})\,.
\end{eqnarray}

\end{widetext}

\section*{Acknowledgments}

DB thanks Chris Kavanagh for useful discussions.
DB also thanks ICRANet and the italian INFN for partial support and IHES for warm hospitality at various stages during the development of the present project.

\end{document}